\def \Lya{{Ly-$\alpha$}}

\def \cf{cf.}
\def \eg {{e.g., }}

\def \etal{{et~al.}}
\def \ie{{i.e.,}}

\def \kms{km~s$^{-1}$}
\def \apj{ApJ}
\def \apjs{ApJS}

\documentstyle[11pt,aaspp4,tighten]{article}
\begin{document}
\input psfig
\footnotesep=7pt

\title{THE HST QUASAR ABSORPTION LINE KEY PROJECT. XIV. \\ The
Evolution of \Lya\ Absorption Lines in the Redshift\\ Interval 0 to
1.5 \altaffilmark{1} \altaffiltext{1}{Based on observations with the
NASA/ESA {\it Hubble Space Telescope}, obtained at the Space Telescope
Science Institute, which is operated by the Association of
Universities for Research in Astronomy, Inc., under NASA contract
NAS5-26555.} }

\author{
Ray~J.~Weymann,\altaffilmark{2} \altaffiltext{2}{The Observatories of
the Carnegie Institution of Washington, 813 Santa Barbara Street,
Pasadena, CA 91101, email:~rjw@ociw.edu}
Buell~T.~Jannuzi,\altaffilmark{3} \altaffiltext{3}{National Optical
Astronomy Observatories, Science Operations Division, P.O. Box 26732,
Tucson, AZ~85719, 
email:~jannuzi@noao.edu}
Limin Lu,\altaffilmark{4} \altaffiltext{4}{Department of Astronomy,
Robinson Laboratory, California Institute of Technology,
Pasadena, CA 91125}
John~N.~Bahcall,\altaffilmark{5}\altaffiltext{5}{Institute for Advanced 
Study, School of Natural Sciences, Olden Lane, Princeton, NJ~08540,
email:~jnb@IAS.edu, sofia@IAS.edu} 
Jacqueline~Bergeron,\altaffilmark{6}\altaffiltext{6}{European Southern 
Observatory, K. Schwarzschild--Str. 2, D--85748, Garching \\ GERMANY,
email:~jbergero@eso.org} 
Alec~Boksenberg,\altaffilmark{7}\altaffiltext{7}{Institute of
Astronomy, University of Cambridge, Madingley Road, Cambridge CB3 OHA,
UK, email:~boksy@ast.cam.ac.uk }
George~F.~Hartig,\altaffilmark{8}\altaffiltext{8}{Space Telescope
Science Institute, 3700~San Martin Drive, Baltimore, MD~21218,
email:~hartig@stsci.edu}
Sofia~Kirhakos,\altaffilmark{5}
W.~L.~W.~Sargent,\altaffilmark{9}\altaffiltext{9}{Robinson Laboratory
105-24, California Institute of Technology, Pasadena, CA 91125,
email:~wws@deimos.caltech.edu }
Blair~D.~Savage,\altaffilmark{10}\altaffiltext{10}{Department of Astronomy, 
University of Wisconsin, 475 N. Charter Street, Madison, WI 53706,
email:~savage@uwast.astro.wisc.edu } 
Donald~P.~Schneider,\altaffilmark{11}\altaffiltext{11}{Department of
Astronomy and Astrophysics, The Pennsylvania State University,
University Park, PA 16802, email:~dps@astro.psu.edu}
David~A.~Turnshek,\altaffilmark{12}\altaffiltext{12}{Department of
Physics \& Astronomy, University of Pittsburgh, Pittsburgh, PA 15260,
email:~turnshek@vms.cis.pitt.edu }
\& Arthur~M.~Wolfe\altaffilmark{13}\altaffiltext{13}{Center for
Astrophysics \& Space Sciences, C011, University of California San
Diego, La Jolla, CA 92093, email:~art@ucsd.edu} 
}
\begin{abstract}
We present the results of an analysis of the rate of evolution of the
\Lya\ absorption lines in the redshift interval 0.0 to $\sim$1.5 based
upon a sample of 987 \Lya\ absorption lines identified in the spectra
of 63 QSOs obtained with the Faint Object Spectrograph (FOS) of the
{\it Hubble Space Telescope} (HST). These spectra were obtained as
part of the QSO Absorption Line Survey, an HST Key Project during the
first four years of observations with the telescope. Fits to the
evolution of the number of absorbers per unit redshift ($dN/dz$) of
the form $ dN/dz = A \times (1+z)^{\gamma}$ continue to yield values
of $\gamma$ in the range 0.1 to 0.3, decidedly flatter than results
from groundbased data pertaining to the redshift range $z > 1.7$.
These results are consistent with our previous results based on a much
smaller sample of lines, but the uncertainties in the fit have been
greatly reduced. The combination of the HST and groundbased data
suggest a marked transition in the rate of evolution of the
\Lya\ lines at a redshift of about 1.7.  The 19 \Lya\ lines from an
additional higher redshift QSO from our sample for which tentative
line identifications are available (UM~18, $z_{\rm em}=1.89$), support
the suggestion of a rapid increase at around this redshift.

We derive the cumulative distribution of the full sample of 
\Lya\ lines and show that the distribution in redshift can indeed be
well represented by a power law of the form $(1+z)^{\gamma}$.  For
this same sample, the distribution of equivalent widths of the
\Lya~absorbers above a rest equivalent width of 0.1~\AA~is fit
quite well by an exponential.

Comparing samples of \Lya~lines, one set of which has redshifts the
same as, or very near to, the redshifts of ions from heavy elements,
with another set in which no ions from heavy elements have been
identified, we find that the \Lya~systems with heavy element
detections have a much steeper slope than the high rest equivalent
width portion of the Lyman--only sample. We argue that this result is
not likely to be due to either line misidentification or incomplete
spectral coverage. Considering the insensitivity of the equivalent
width to large changes in the column density for saturated lines, we
suggest that this result is probably attributable to rapid evolution
of the very highest column density systems, rather than real
differences in metallicity.  We find evidence that the rate of
evolution increases with increasing equivalent width.

We compare our results for the variation of line density with redshift
to recent numerical simulations of \Lya\ absorbers, in particular, to
those of \cite{riedig98} which extend to zero redshift. We find fairly
good agreement between these simulations and our results though the
rapid evolution we find in the \Lya\ systems containing heavy element
ions is not predicted in their models. We speculate that these heavy
element containing \Lya\ systems involve those clouds closely
associated with galaxies,
whose column densities are too high and whose sizes are too small to
be included in the Riediger \etal\ simulations.  

Our results for \Lya\ lines at the high end of our equivalent width
distribution are compatible with the recent analysis of the
absorber--galaxy correlation by \cite{chen98}. For the weaker lines
however, our results suggest that whatever association exists between
absorbers and galaxies is different from that for the stronger lines.
We conclude with some suggestions for further observations.
\end{abstract}


\section{Introduction} 
\label{introduction}

The Hubble Space Telescope QSO Absorption Line Survey, a Key Project
during the first four years of observations with the telescope, has
resulted in the acquisition of ultraviolet spectra of a sample of 92
bright QSOs. These observations have yielded valuable information on
the properties of absorption systems at low redshift. In particular,
they have allowed preliminary estimates to be made of the rate of
evolution of the \Lya~absorption systems. Discussions of our past
results based on analysis of about 10\% of the \Lya~absorbers in our
data set can be found in our previous two ``catalog papers''
(\cite{bahcall93,bahcall96}; hereafter CAT1 and CAT2).  We have
recently completed a significant expansion of the database available
for analysis of the evolution of the \Lya\ lines both in terms of the
total number of lines available and in the range of redshift covered
by the sample.  This new database is described in the third of the
catalog papers (\cite{jannuzi98a}; hereafter CAT3). We refer hereafter
to the data set resulting from these 3 papers as the Catalog.  This
new database makes it appropriate to describe in more detail the
results, methodology, and limitations of our analysis of the evolution
of the low redshift \Lya\ absorption systems, which is the purpose of
the present paper.

In two separate publications we will discuss the correlation
properties of the \Lya\ absorption systems (\cite{jannuzi98b}) and the
inferences concerning the ionizing background radiation which can be
drawn through the ``proximity effect'' (\cite{lu98a}).

During the past few years there has been a maturing of attempts to
understand the nature of the \Lya\ absorption clouds, evolving from
rather simple idealized models (e.g. pressure--confined clouds,
gravitationally confined ``mini--halos'', freely expanding clouds,
caustics etc.) to sophisticated numerical simulations which are able
to reproduce many of the features of the \Lya\ absorption in the
context of the evolution of structure. 

The spatial and mass resolution in current simulations is still too
coarse and the processes involving star formation and the details of
galaxy formation and evolution still too poorly understood to allow
detailed predictions of absorbers on sub--galactic scales within
galaxies.  Moreover, simulations incorporating hydrodynamics are still
so time-consuming that such simulations are just now being extended to
zero redshifts.  However, we are aware of one simulation which has
been carried out to a redshift of zero (but without a full
hydrodynamic treatment) at the level of detail and resolution so as to
allow useful comparisons to be made between the data and the
simulations (\cite{riedig98}.) Continued rapid increases in computing
power should allow detailed comparisons between full hydrodynamic
simulations and observations near zero redshift in the near future. (A
comprehensive list of references to simulations and a discussion of
some of the issues involved in the simulations may be found in
\cite{riedig98}.)

Among the obvious and crucial areas for comparison are the
distributions of line density as a function of redshift and rest
equivalent width, and indeed this was one of the main initial goals of
the QSO Absorption Line Key Project.  We hope that the present
results, in conjunction with the wealth of groundbased data at higher
redshifts, will serve as a starting point for such detailed
comparisons.

The organization of the remainder of this paper is as follows: In \S2
we describe the subset of QSOs taken from the Catalog which are used
in the present analysis. In \S3 we discuss the applicability of ``line
counting'' and maximum likelihood techniques to the analysis of the
evolution of the \Lya~lines and describe the various parameters
characterizing subsamples of the \Lya\ lines used in our analyses.  In
\S4 we present the results for the evolutionary rates for these
samples. We discuss in \S5 the distribution function of the \Lya\
lines as a function of redshift and of rest equivalent width
(hereafter REW).  We also briefly discuss the corresponding
distribution for $N_{HI}$ column densities for an assumed Doppler
parameter. We give in \S6 an alternative depiction of the evolution of
the \Lya\ lines by binning the data, and discuss the effect of adding
lines from the single high redshift ($z \, > \, 1.5$) QSO (UM~18,
$z_{\rm em}=1.89$) for which we have a complete, though tentative, set
of line identifications. In \S7, we ask whether there are differences
in the evolutionary rate for \Lya\ systems having, and not having,
detections of lines which can be identified with heavy elements and
investigate whether there is a dependence of the rate of evolution
upon the strength of the \Lya\ lines. We conclude in \S8 with a
summary of the results, a comparison of our results with the numerical
simulations of \cite{riedig98}, some implications of our results
concerning the nature of absorber--galaxy associations, some
speculations on multi--component \Lya\ models and suggestions for
further observations. Some details of the maximum likelihood code used
in our analysis are described in the Appendix.

\section{The QSO Sample} 
 
In CAT3 we present all previously unpublished spectra taken as part of
the HST QSO Absorption Line Survey. However, for eight of the QSOs
(all with $z>1.0$, and including the highest redshift objects in the
sample), while we were able to measure wavelengths and equivalent
widths for the absorption lines, we have not yet produced a
significantly complete list of line identifications based upon the
rigorous identification algorithms presented in CAT1, CAT2, and
CAT3. Because unambiguous identifications become increasingly
difficult for the increasing line density of the higher redshift
objects, and because the decomposition of blended features in a unique
way also becomes very difficult at these redshifts, especially at the
relatively low resolution of our spectra (230--270 \kms), it is not
clear if it will be possible to produce completely and securely
identified line lists for this set of objects based upon our current
data.  For a ninth high redshift object, UM~18 ($z_{\rm em}=1.89$) we
have managed to complete a set of identifications for the measured
lines, but for the reasons just stated pertaining to the eight objects
mentioned above, we consider these identifications to be less robust
than the identifications of the lines in the remainder of the
Catalogue. For the purposes of this paper we consider the
identifications for UM~18 to be tentative (see CAT3 for additional
discussion of the line identifications for UM~18).

For these reasons we have decided to summarize our conclusions
concerning the evolutionary properties of the \Lya\ lines based upon
the completely identified line lists published in the Catalog, not
including the UM~18 line list for the majority of the analysis.
However, we do briefly describe the impact of adding this higher
redshift quasar on the results.

Illustrations of all the spectra, the variation of the observed
equivalent width of the line detection threshold as a function of
wavelength, and the detailed line lists are all found in the Catalog.
Note that in the present paper we only consider lines in the
``complete sample'', which is defined to be those lines with a
significance level, $SL$, of at least 4.5 (\cite{schneider93},
hereafter Paper~II; CAT1; CAT3).  The significance level is defined as
follows:
\begin{equation}
SL~=~{\vert W \vert \over {\overline \sigma}(W)}~~~, 
\label{eq:siglevel}
\end{equation}
where $W$ is the observed equivalent width, and ${\overline
\sigma}(W)$ is the 1$\sigma$~error in the observed equivalent width of
an unresolved line.  Note that ${\overline \sigma}(W)$ is calculated
with the flux errors at the positions of strong absorption lines
replaced by the errors interpolated from the surrounding continuum
points (see Paper~II).  The significance level differs from the more
familiar definition of signal-to-noise ratio, $SNR = | W | /
\sigma(W)$, because of the use of the {\it interpolated} error array
for the calculation of~$SL$.  

As noted above, there are several objects in the Catalog for which
there are line lists, but not secure line identifications. In
addition, there are a few objects which have deep broad absorption
troughs (BALQSOs) and which have been omitted in the analysis of this
paper; these objects will be described elsewhere (\cite{turnshek98}).
Finally, there are a few objects where the spectral coverage was such
that no region shortward of the \Lya~emission line redshift was
observed. These objects too, are of course omitted from the present
analysis, as are all the objects which were only observed using the
low dispersion G160L grating, which has a resolution of $\approx$ 1100
km~s$^{-1}$.

There remain 63 QSOs (not counting UM~18) whose \Lya~lines comprise
the data set analyzed here.  In Table~1 we list these objects in order
of emission line redshift. Column 1 lists the common name, columns 2
and 3 the 1950 coordinates, column 4 the emission line redshift, and
columns 5 and 6 the lower and upper usable redshifts for detecting a
\Lya\ absorption line. The lower usable redshift is defined by the low
wavelength cutoff of the spectral coverage, (except that wavelengths
shortward of 1218~\AA~were not considered.)  The upper usable
redshift coverage was set by the smaller of the upper wavelength
cutoff of the spectral coverage and the wavelength corresponding to
the ``proximity velocity '' cutoff, discussed in~\S3. For Table~1 this
cutoff velocity was chosen to be 3000 km~s$^{-1}$.  Additional
information about each of the QSOs listed in Table~1 can be found in
Table~1 of CAT3.

\section{Application of Maximum Likelihood Analysis to Line Counts}
\subsection{Is Line Counting Appropriate?}
In this section we describe the use of maximum likelihood estimators
to obtain values of three parameters which characterize the
distribution of the \Lya~absorption lines as a function of REW and
redshift. From the point of view of the physical properties of the
absorbing gas it would, of course, be preferable to know the
distribution of Doppler widths and H~I column densities, but at our
resolution the lines are not resolved, so profile fitting is not
feasible. Although use of the higher Lyman lines and curve-of-growth
techniques would allow us to infer something about Doppler widths and
column densities for systems in which the higher Lyman lines are
observed, this subset of the data is substantially smaller than for
\Lya~itself. Moreover, blending of components could very well vitiate
these results.

Additionally, at the observed line densities at the highest redshifts
for objects in CAT3, not only do line identifications become more and
more difficult, but, at the FOS resolution ($\sim$230--270 km
s$^{-1}$), line blending becomes more and more severe, and the
decision about when to decompose a given feature into several
components becomes increasingly arbitrary---though reproducible---in
terms of well--defined algorithms and reduction parameters
(Paper~II). More generally, some simulations call into
question whether the concept of a ``cloud'' has much meaning. In this
sense, it would be worthwhile to ultimately repeat our entire analysis
based upon the flux--deficit method, as explicated in detail in
\cite{press93}. In both approaches the most difficult and least
objective procedure still remains the drawing of the continuum.

Notwithstanding these shortcomings and the fact that we must deal
with equivalent widths rather than column densities, we take the point of view that
simulated spectra can be reduced and analyzed using the same
techniques as the real data. Direct comparisons between
the equivalent widths in the real and simulated data can be made and
the  question of line counting
{\it vs.} flux deficit is thus somewhat moot.

Finally we have generally not attempted to evaluate the reliability
of every line identified as \Lya\ in the Catalog, but have taken a
line to be \Lya\ whenever it was the first identification appearing in
the line lists of the Catalog, independent of any indications of
blends or alternative, but less probable, identifications.  The
exception to this statement occurs where candidate C~IV doublets are
identified shortward of the \Lya\ emission line and when the expected
position of the corresponding \Lya\ is not within our spectral
coverage (or falls where the signal-to-noise ratio is very poor) and
when there are no other lines identified at the same redshift as the
candidate C~IV doublet.  In such cases the algorithms used for the line
identifications cannot provide an unambiguous identification
and we had to make a subjective decision in selecting the preferred
identification (see \S3.4 in CAT3).

We have carried out simulations in order to estimate the probability
that such C~IV doublets are actually chance pairs of \Lya\ lines which
satisfy the criteria used by the line identification algorithm to
accept a C~IV doublet, and these simulations are described in detail
in CAT2 and CAT3. (A similar ambiguity occurs for the O~VI doublet, but
in practice this has very little impact on the line lists.)

We have treated these relatively small number (46) of ambiguous C~IV
doublets which occur in the line lists for the objects considered in
this paper in 4 different ways: (1) Regard all such possible C~IV
doublets as being two \Lya\ lines. We refer to the line list treated
in this way as the ``maximum \Lya\ line list'' assumption. (2) Regard
all such possible C~IV doublets as actually being C~IV.  We refer to
the line list constructed in this way as the ``minimum \Lya\
line list'' assumption. (3) Based upon the probabilities for finding
false doublets (see Table~4 of CAT3) we assign probabilities to the
identifications of being either C~IV doublets or pairs of \Lya\
lines. Then, with a random number generator, assign each of these
possible C~IV doublets to be either two \Lya\ lines or a C~IV doublet
on the basis of these probabilities. We refer to this case as the
``statistical \Lya~line list''.  (4) We use the line identification
appearing first in the CATALOG and refer to this as the ``preferred
\Lya\ line list''.

\subsection{Parameters Characterizing \Lya\ Line Subsamples}
The maximum likelihood code used in the present paper is based upon
the description given in the paper by \cite{murdoch86} but differs in
a few significant ways. The formalism and these differences are
described in detail in the Appendix. The explicit parameterization
adopted to describe the evolution is given in equations A-1, A-2, and
A-3 of the Appendix. We describe there also the determination of the
distribution functions in redshift and rest equivalent width and the
corrections for incompleteness.

In this section we define various ways of dividing the \Lya\ lines
into subsamples. The results will be presented and discussed in \S4.
Most of our results are contained and summarized in a table (Table~2)
in which each row of the table specifies and contains the results for
one of the subsamples.  The columns of Table~2 have the following
meaning: Column 1 is the number identifying the sample. Columns 2-6
define the particular sample of \Lya~lines: Column 2 gives the number
of lines in the sample, column 3, (labeled REW) gives both the
limiting lower and upper rest equivalent widths accepted in that
sample. The notation L,M,B in column 4 indicates whether the samples
included: i) \Lya\ lines in which there were no other lines identified
with heavy elements at or near the redshift of these \Lya\ lines (L);
(ii) {\it only} such systems associated with heavy--element containing
systems (M), or (iii) both types (B).  By ``associated'' we mean that
the \Lya~system in question lies within 300 km~s$^{-1}$ of a system
(most often the same system) containing at least one heavy element
ion.\footnote{Note that the use here of the phrase ``associated'' is
different from that sometimes used to denote absorption systems whose
redshift is very close to the emission line redshift.}

In column 5 a ``V'' (for variable sensitivity) indicates that
\Lya~lines were accepted regardless of the value of the 4.5 $\sigma$
REW detection limit at the wavelength in question (of course the line
itself must exceed this limit to be accepted in the sample). As
explained in the Appendix, the incompleteness in the ``V'' sample is
corrected for by the fit to the exponential distribution of rest
equivalent widths. Thus, the validity of the incompleteness correction
in the ``V'' samples depends explicitly on the goodness of fit to an
exponential in the distribution of rest equivalent
widths. Alternatively, we also define a ``U'' sample (for uniform
sensitivity), accepting only those lines whose REW is greater than
some minimum value, but also limiting the spectral coverage in which
lines are accepted, regardless of their REW, to regions having at
least this minimum detection limit.  In \S4 we choose the value of
this minimum detection limit in order to maximize the number of lines
in the large ``U'' sample. We stress that the purpose of defining such
U samples is simply to check that the values of $\gamma$ we find for
the V samples are not strongly affected by the assumption of an
exponential distribution of rest equivalent widths. For the U samples,
the determination of the best fit for $\gamma$ is independent of any
assumption about the distribution of equivalent widths above the
adopted minimum detection limit for the U sample. However this
independence is achieved at the cost of significantly decreasing both
the number of lines and the spectral coverage.  Column 6 is the
``proximity value'' cutoff: Using the emission line redshifts given in
Table~1, we calculate the wavelength corresponding to a Doppler shift
of $V_{prox}$ from the emission line redshift, where $V_{prox}$ is
taken positive for wavelengths shorter than the \Lya~emission
line. There are several reasons for treating with caution lines near
the emission line redshift (which is itself uncertain when it is not
based upon lines from the narrow emission line region, as is the case
for many of the QSOs in Table~1). First, the ionization of the clouds
may be influenced by the radiation field from the QSO itself, and
indeed quantitative analysis of this effect is now a standard tool to
use in estimating the metagalactic ionizing radiation field, as is
discussed for the Key Project sample by \cite{lu98a}.  However, there
are at least two other equally important reasons: (1) In the region
surrounding a QSO, there is likely to be enhanced density, with an
enhanced density of \Lya\ lines, mitigating to some extent the
ionizing effect of the QSO (\cf~\cite{bechtold94,williger94}) (2)
Recent work---\cf~articles in a recent PASP Conference Series
(\cite{barlow97,hamann97,aldcroft97})--- clearly demonstrates that a
significant fraction of such lines---perhaps the majority---are
intrinsic to the QSO. Such absorption features---originally called
``associated absorption'' (\cf~\cite{foltz88})--- typically have very
high levels of ionization, and they are seen frequently in O~VI and
C~IV in the key project sample (\cf~CAT3).  Generally, their ejection
velocities do not exceed 3000 km~s$^{-1}$, and for this reason we have
adopted this value in all the samples in Table~2 except for sample
1. However, it is also now clear (\cite{hamann97}) that there are
occasional instances of material ejected to very high
velocities---presumably related to the Broad Absorption Line QSOs
(BALQSOs). In any particular instance it is not obvious whether a
given absorption feature is truly intervening or ejected, and indeed,
high ionization troughs in one of the objects in our sample,
PG~2302+029, is such an instance, as recently discussed by
\cite{jannuzi96}.

Columns 7, 8 and 9 contain the results and are, respectively, the line
densities at $ z \,= \, 0$, the slope, $\gamma$, and the
characteristic rest equivalent width in the exponential distribution,
$w^*$. Also given are the 1$\sigma$ uncertainties in these three
quantities, calculated according to the algorithm discussed in the
Appendix. Note that in column 7, where we give the line density at
zero redshift, we give the line density for lines above a rest
equivalent width of 0.24~\AA, using the value of $w^*$ given in column
9. Line densities normalized to any other lower limiting REW can be
readily calculated by multiplying by $exp[-(w_{lim} -0.24)/w^*]$.

\section{Results for Several Samples} 

In this section we define several subsamples of the \Lya\ lines and
evaluate the maximum likelihood parameters for each. However, as
discussed in \S3, there is some ambiguity in the identification of a
subset of lines which could be identified as either pairs of \Lya\
lines or a C~IV doublet. We defined in \S3 four possible sets of
identifications which we characterized as the ``minimum'',
``maximum'', ``statistical'' and ``preferred'' \Lya\ line lists. As
described below in connection with sample~2, we conclude that the
results are relatively insensitive to which of these line lists we
choose.

The first sample, sample~1 in Table~2, presents the results using the
preferred line list. For each of the spectra of the 63 QSOs of
Table~1, (except for the \Lya~lines identified in the high redshift
object UM~18, which are not included and are discussed in \S6 below)
sample~1 includes all \Lya\ lines in the preferred list whose
absorption redshifts are less than the emission line redshifts listed
in Table~1.  Sample~2 shows the results of eliminating those lines
within 3000 km~s$^{-1}$ of the emission line redshift: because there
are relatively few such lines, their inclusion or exclusion does not
significantly alter the results of these large samples.

In order to check the sensitivity of our results to the uncertainties
in the line identifications noted above, we considered the same
criteria for line selection used in sample~2, but used the minimum,
maximum, and statistical line lists.  The results for $w^*$ and the
line density differ only very slightly. The results for the minimum
line list yield a value of $\gamma$ of 0.08, slightly flatter than for
the preferred list, but by an amount which is only one half of the
formal 1$\sigma$ uncertainty in the fit. In the case of the minimum
line list, for any given line of sight, \Lya\ lines in the preferred
line list, whose identifications are ambiguous (in the sense of \Lya\
pairs vs. a C~IV doublet) are removed at the low redshift range
covered by that line of sight, since the ambiguity is significant only
when the \Lya\ line which would normally be required to confirm the
C~IV identification is at too low a wavelength to be observed at all,
or is observed with very low signal to noise.  This effect therefore
tends to steepen $\gamma$. However the QSOs in whose lines of sight
these ambiguities tend to occur most frequently are those at the
higher redshifts, and thus, considering the entire sample, it is the
higher redshift lines which tend to be removed from the preferred
sample, which flattens $\gamma$. For the maximum line list, the results
are in the opposite sense and we obtain a $\gamma$ of 0.24, again
differing from the preferred line list value by only one half of the
formal 1$\sigma$ uncertainty in the fit. Finally, the ``statistical''
line list gives a result which is close to, but slightly less than,
the $\gamma$ for the maximum line list.  Since all of these line lists
yield values of $\gamma$ which do not differ by more than one half
$\sigma$ from the formal uncertainty in the fit of $\gamma$, in the
remainder of this paper we use the preferred line list.

In \S3 we described the use of a uniform detection limit sample, which
avoids any assumption about the distribution of rest equivalent
widths. Sample~3 differs from sample~2 in utilizing such a uniform
detection limit sample and shows that use of a uniform sensitivity
sample also does not substantially alter the results for $\gamma$ for
these larger samples.  For sample~3 we chose the value of the minimum
REW to be 0.24~\AA, which maximizes the number of lines in the sample:
If we take a value larger than this, we expand the wavelength coverage
but restrict the number of lines, since their density decreases
rapidly with increasing REW. Conversely, if we lower this minimum REW
further, we restrict further the wavelength regions having this more
sensitive detection limit, and thereby accept fewer lines.  In
sample~3 we have therefore set the lower limit of the REW for lines
accepted in the sample to be 0.24~\AA~and thus also to regions of the
spectra having at least this sensitivity. Since there are
significantly fewer lines, and more restricted wavelength coverage,
the formal errors are larger than in sample~2.

As discussed in \S7 it has been traditional to separate the \Lya~lines
into ``Lyman--only'' systems and those containing at least one heavy
element line (or ``associated'' with such a heavy element--containing
system in the sense defined above). Samples~4 and~5 are for V and U
samples comprising just the Lyman--only systems. Once again, due to
the fact that such lines are in the heavy majority, the results do not
differ much from sample~2.

The most striking thing about the results in the first five samples of
Table~2 are the very low values of $\gamma$ which are much lower than
was initially expected, on the basis of groundbased samples, when the
Key Project quasar absorption line survey was conceived. These low
values were suggested in our preliminary analysis in both CAT1 and
CAT2, but are now even smaller, are based upon a much larger number of
lines, and now extend to redshifts of 1.5. Similar preliminary
results, based upon spectra of two QSOs of lower resolution and a very
much smaller sample of lines, were obtained by \cite{impey96}. More
recently, a preliminary analysis based upon archival FOS-HST data has
been presented by \cite{dobrbech97} with similar results.  At some
redshift, however, we certainly expect an upturn in the rate of
evolution if these results are to be compatible with the groundbased
results. We return to this topic in \S6.

The slow rate of evolution we observe seems in conflict with the
visual impression resulting from inspecting the spectra of the QSOs at
$z\approx0.3$ and $z\approx1.0$ (see figures in CAT1, CAT2, and CAT3).
The $z_{\rm em} \ge 1.0$ QSOs certainly appear to have many more
strong absorption lines blueward of the QSO Ly-$\alpha$ emission line
than does, for example, the spectrum of the low redshift QSO
3C~273. Three factors contribute to this impression.  First, while the
spectra obtained in the survey are of fairly uniform signal--to--noise
ratio per resolution element (about 30) over most of the wavelength
coverage, this results in a lower {\it rest} equivalent width
detection limit in the spectra of the higher redshift objects than it
does in the lower redshift objects.  Moreover, redshifts from $\sim$0
to about 0.35 are seen only in the G130H spectra, which sometimes have
somewhat lower S/N than the G190H and G270H data, especially at the
lowest redshifts. This results in correspondingly fewer lines at low
redshift.  Second, there are more heavy element systems detected in
the spectra of the higher redshift objects (in part because of a
larger number per unit redshift, but also because there is more
observed path length at the higher redshifts) and each of these
systems can contribute 3 to 20 absorption lines to the
spectrum. Finally, the higher order Lyman series lines (Ly-$\beta$,
Ly-$\gamma$, etc.)  are more readily observable in the spectra of the
higher redshift QSOs, due to a larger observed path length and the
more sensitive rest equivalent width detection threshold.  All of
these effects are evident in the spectrum of the $z_{\rm em}=0.770$
QSO PG~1538$+$477 (see CAT3 and figure 3 of \cite{jannuzi98}).

We illustrate the rest equivalent width detection limits in Figure~1,
where we reproduce from CAT3 both the {\it observed} 4.5$\sigma$
detection limit and the {\it rest} 4.5$\sigma$ detection limit for
typical spectra resulting from the 3 gratings used here. At a redshift
of 1.5, even for observed equivalent width limits which are
comparable, the minimum detectable rest equivalent width is lower by a
factor of 2.5 than it is at zero redshift. Because of the fairly rapid
decline in the distribution of \Lya~lines with increasing rest
equivalent width, these numerous weaker lines are not detectable at
the lower redshifts.  All of these effects combine to give one the
false visual impression of a rapidly evolving population of \Lya~
lines.

\section{The Distribution of \Lya~lines in Redshift, Rest Equivalent
Width and Column Density}
\nobreak
\subsection{The Cumulative Distribution Versus Redshift}

The values of $\gamma$ and $w^*$ summarized in Table~2 are based upon
the assumption that the distribution of the \Lya\ lines can be
reasonably well represented by a power law in $(1+z)$ and an
exponential in the rest equivalent width.  Since the effective path
length in redshift space is affected both by the number of lines of
sight associated with each grating, the distribution of emission line
redshifts in our sample, and the variation in minimum detectable
equivalent width as a function of wavelength for a given grating,
there are significant differences between the observed redshift
distribution of the \Lya~lines and the ``true'' distribution (\ie~the
observed distribution corrected for the non--uniform coverage in
sensitivity and wavelength.)  The same is true for the distribution of
rest equivalent widths. The observed distribution of lines in redshift
and REW is shown in Figure~2, where we have also plotted the number of
sightlines as a function of redshift.

In the Appendix, we describe the algorithms used for making these
corrections. In displaying the results we could, following for example
\cite{storrie96}, plot the cumulative  distribution {\it vs.}
the ``effective path length'', in which the cumulative distribution
function increases by an equal amount for every
\Lya~line. Alternatively, we could weight the lines in forming the
cumulative distribution function, as described in the Appendix, and
plot this cumulative distribution directly as a function of
redshift. The former has the advantage that a statistical test for
goodness of fit ($e.g.,$ the KS test) can be rigorously applied,
whereas the latter has the advantage that ``redshift'' is much more
readily interpretable than ``effective path length'' as the
abscissa. We opt for the latter case. Where the data are sparse
(e.g., between the redshifts of 0.31 and 0.35 corresponding to the
region from 1600~to 1650~\AA~where the S/N is generally very
poor---see Figure~1) each line is weighted more heavily than in
adjacent regions and causes a jump in the ordinate of the distribution
function which is larger than for a line in a well--sampled redshift
interval.

Nevertheless, as shown in Figure~3, it is evident that the power law
fit to the data of sample~1 is a remarkably good one.  The application
of the KS test to the data displayed in Figure~3, while not rigorous
for the reasons discussed in the preceding paragraph, yields a
probability of~36\% that a KS statistic $\geq$ to the observed one
would arise from a random sampling of true power law distributions
with the same slope, thus confirming the visual impression of the
goodness of fit.  Since it is rather unlikely that the evolution of
such complex physical processes as those which produce \Lya~lines in
QSO spectra should be represented accurately by a power law, we do not
pay particular attention to whether a better representation might be
found---as always, our point of view is that the empirical (but
corrected) distribution function is to be compared with detailed
numerical simulations.

\subsection{Distribution of Equivalent Widths}
A similar analysis can be carried out for the distribution of
equivalent widths. In this case, it is more instructive to compare the
raw and corrected {\it differential} distributions in bins of
0.1~\AA. This is shown in Figure~4. The fit is quite good. There may
be a very slight excess of lines in the first bin, though this is not
statistically significant. The fact that the exponential is a fairly
good fit follows from the convolution of a power law in the
distribution of H~I column densities over a range of Doppler
parameters, as pointed out by \cite{murdoch86} and \cite{press93}.

Since most simulations are presented in terms of the H~I column
density, however, we make a very rough attempt to translate the
results of Figure~4 from equivalent width to column density. This
requires knowledge of the Doppler parameters which we lack. There is
to date only limited information on the distribution of Doppler
parameters at low redshifts, but the existing data suggest it is not
very different from that found for higher redshift systems.  In
Figure~5 we convert equivalent widths to column densities for a single
value of the Doppler parameter of 30 km~s$^{-1}$, though there are
clearly instances of lines having both larger as well as smaller
values.  The approximate slope of the differential column density
distribution, $dlog N/dlogN_{HI}$ is~$-1.3$~at the low column density
end, consistent with values found from high resolution studies at high
redshifts (\cf~\cite{kim97}).  In Figure~5 we also plot the
curve--of--growth for \Lya~relating equivalent width and column
density for 3 Doppler parameters (20, 30, and 40 km~s$^{-1}$), and
from these curves the influence of the Doppler parameter upon the
inferred column density distribution can be estimated.

Thus we conclude that the large samples are quite well represented by
a power law in redshift and an exponential in rest equivalent
width. There remains, however, the question of whether the rate of
evolution might vary as a function of line strength or metallicity: we
defer discussion of this question until \S7.

\section{Is There Evidence for an Upturn in the Rate of Evolution with
Increasing Redshift?} 
As noted in \S4, at some point we should expect that an upturn should
exist in the rate of evolution of the \Lya~lines deduced from the Key
Project data if they are to match smoothly onto the results obtained
from groundbased spectroscopy. There was some slight hint of this from
the data based upon the smaller and lower redshift samples comprising
CAT1 and CAT2. However, when we bin the data from the substantially
larger database represented by sample~2, there is no convincing
evidence for this up to a redshift of 1.5, as shown in Figure~6. In
Figure~6 we have divided the redshift space between 0 and 1.5 into
five bins spaced equally in $\log (1+z)$. Within each bin, we
evaluated $dN/dz$ at the middle of that bin, fixing both $\gamma$ and
$w^*$ to have the values for the global fit of sample~2. (The
resulting values of $dN/dz$ are very insensitive to the value of
$\gamma$ assumed since the values of $dN/dz$ plotted in Figure~6 are
centered between the limits of each bin.) Figure~6 confirms the
impression of a very flat rate of evolution over the range $z$ = 0.0
to 1.5 suggested by Table~2.  The same procedure using the U sample
(sample~5) gives very similar results.  In Figure~6 we have also
plotted groundbased results from an early analysis by
\cite{lu91} and more recent results by \cite{bechtold94}, based upon
data which is similar to ours in terms of line density per resolution
element in the region where the groundbased and HST data join. The
slope found in that analysis was 1.85. (A recent analysis based upon
much higher resolution and signal to noise spectra using
\Lya\ lines above redshift two, and whose strengths are comparable to
those in our sample yielded a steeper slope, $\gamma =
2.78$; \cite{kim97}).  Taken at face value, Figure~6 suggests that 
in the redshift range from about 1.3 to 1.7, which bridges the gap
between our FOS data and the lowest redshifts available from the
groundbased data, the rate of evolution increases rather abruptly.

\subsection{The Addition of UM~18 to the Sample} 
While both the character of the data and the analysis in the Bechtold
data are not very different from our HST data and analysis, it would
be preferable to have a uniform data set and analysis which spans this
redshift range. In particular, adding \Lya\ lines in the redshift
range between $z\sim1.5$ and $z\sim1.7$ from the Key Project objects
is very important, as it will fill in the gap between our currently
well--identified sample of absorbers and the large groundbased samples
that have been compiled.

After the bulk of the analysis for this paper had been performed,
identifications of the lines in the spectrum of one of these objects,
UM~18 ($z_{\rm em}=1.89$) were completed (CAT3). Since line blending
and the location of the continuum start to pose significant problems
in objects at these redshifts, at the FOS G270H resolution, we must
regard the line lists from UM18 as considerably more uncertain than
those for the other 63 QSOs we have discussed. Therefore, we have
chosen not to include UM~18 in most of the analysis. We have, however,
run a sample identical to sample~2 in which all the UM~18 \Lya~lines
are included. Although UM~18 has a large number of lines at redshifts
below $z$=1.5, the overwhelming number below this redshift still come
from other lines of sight, and the inclusion or non-inclusion of UM~18
lines at redshifts below this redshift makes no significant difference
in either the global fits in Table 2 or the binned data of Figure~6.

However, when we extend the binned data to include the bin from 1.5 to
1.69 (corresponding to 3270~\AA, the red end of the G270H spectral
coverage) then the 19 lines contributed by UM~18 in this region
suggest a sharp upturn in the rate of evolution. The bin spanning this
redshift range and containing these 19 lines is also shown in
Figure~6.  Because of the small number of lines in this bin the upturn
in the line density from what extrapolation of the data up to redshift
1.5 suggests is only significant at about the 1.5$\sigma$ level.
While it is thus somewhat reassuring that there is a moderately strong
indication from UM 18 that we are starting to see this upturn in 
our Key Project data, it will be very important to see if this upturn
is indeed as sharp as suggested from the current data, as the
remainder of the large number of lines in the CAT3 line lists become
reliably identified.  Evidently, the redshift regime between about 1.5
and 1.8 is a very interesting one for further study with both HST and
with groundbased telescopes.

\section{Does Evolution Rate Depend on Metallicity or Line Strength?}
Given the wide range of physical conditions and processes which can
give rise to observable \Lya\ absorption, it would be somewhat
surprising if the rate of evolution did not depend to some extent on
various properties of the \Lya\ absorption such as degree of
association with galaxies, metal content, and line strength.  In this
section we discuss the latter two possibilities.

\subsection{Does Evolution Rate Depend Upon the Presence of Heavy Elements?}
It has been customary to treat the \Lya~absorption systems in which no
other ions are found separately from those in which ions from heavy
elements are detected. Historically, this arose from the original
suggestion of \cite{sybt80} that the low column density Lyman--only
systems represented unprocessed material and formed a population
distinct from absorption systems containing, e.g. C~IV. This idea was
supported by the apparently quite different clustering properties of
the two types of systems.

The advent of very high resolution, high sensitivity groundbased
spectra has cast some doubt on this point of view, since C~IV systems
seem to generally be found at H~I column densities down to $\sim3
\times 10^{14}$ (\cite{cowie95}) below which it has been presumed that
the C~IV lines become too weak to detect individually.  It has further
been suggested (\cite{cowie95}) that there is a ``floor'' in the
metallicity of the \Lya~forest lines of order 10$^{-2}$ solar,
reflecting perhaps general contamination from the first epoch of star
formation.  On the basis of the coaddition of a large number of weaker
Lyman lines (in the H~I column density range $10^{13.5}-10^{14}$)
though, it was found that the ratio of C~IV to \Lya\ strengths become
very much smaller, either because of very much lower (or zero)
metallicity, or conceivably because of an abrupt change in the level
of ionization (\cite{lu98b}). However, using a different technique,
Cowie and Songaila (\cite{cowsong98}) find that the metallicity (as
judged by the C~IV/\Lya~line ratios) does {\it not} drop at very low
column densities (and thus in regions of nearly zero over-density).

The foregoing all refers, of course, to much higher redshift systems
than in our HST sample, and, in view of the rather unexpected flatness
of the rate of evolution of the total ensemble of \Lya~lines (as
reflected in sample~2 of Table~2), it is of interest to ask whether
there is any evidence that the \Lya~systems in which one or more ions
of heavy elements have been detected behave any differently than the
Lyman--only systems, insofar as their rate of evolution is
concerned. We stress that with the sensitivity and limited spectral
coverage of our FOS data, {\it the lack of detection of lines from
heavy elements does not imply that these systems truly lack heavy
elements or are even strongly metal deficient}. Most, perhaps all, may
well be found to contain weak heavy element lines with STIS and/or
groundbased resolution and sensitivity. Thus, although we will
continue to refer to this set as ``Lyman-only'' systems, this should
be understood to mean ``systems with no identified heavy element ions
due to inadequate sensitivity or spectral coverage, or low heavy
element abundance.'' The much smaller set of \Lya\ lines for which
heavy element ions have been identified at the same redshift we
designate as LWHED lines (\Lya\ Lines With Heavy Element
Detections). We also include in this category the very few
\Lya\ lines without heavy element ions identified at the \Lya\ redshift,
but whose redshift is within 300~km~s$^{-1}$ of heavy
element--containing systems, since it seems very likely that such
systems are physically associated with the heavy element--containing
systems.  In defining the LWHED samples it is also crucial to exclude
lines within 3000~km~s$^{-1}$ of the emission line redshift, since,
as remarked previously, systems showing C~IV and O~VI near the
emission line redshift are quite common and have a high probability of
being intrinsic to the QSO.  In our total sample of \Lya~lines with
V$_{prox} =$ 3000 km~s$^{-1}$, there are a total of 78 systems in
which one or more heavy element ions have been identified. There are
an additional five ``associated'' \Lya\ lines (as defined in \S3.2)
for a total of 83 which we place in the LWHED category. Among these
83, however, are some found in the possible BALQSO PG~2302+029 which
has broad weak troughs of O~VI, N~V, C~IV and probably \Lya\, and is
described in detail in \cite{jannuzi96}. Whether these troughs
represent intervening or ejected material is not known, but if the
troughs are interpreted as high velocity ejected material, then, since
sharp components are often found to accompany the broad troughs, for
the purpose of the following discussion we have dropped PG~2302+029
from the samples discussed in this section. There remain 79 LWHED
systems; the five ``associated'' systems and 74 \Lya\ lines having
redshifts identical, within the measuring errors, to the redshifts
defined by the heavy element lines themselves. Although comparison of
the large ``V'' and ``U'' samples~2 and~3 give quite comparable
results, the smaller samples become increasingly vulnerable to the
assumption of exponential fits to the equivalent width
distribution. This is especially true of the LWHED systems, where the
$e$--folding REW, $w^*$, is much larger than in the other samples.  In
samples~6 and~7 therefore, we show the results for both the ``V'' and
``U'' assumptions for the LWHED lines. In the latter case we have
chosen the minimum equivalent width cutoff to maximize the number of
lines, and this occurs at a rest equivalent width of 0.40~\AA. We find
a very much more rapid rate of evolution ($\gamma \sim1.2-1.6$) for
these two samples, respectively, than for the total sample
\hbox{(\eg~sample 2 $\gamma = 0.16$).} Because of the small
number of lines the uncertainty of $\gamma$ for the LWHED samples is
now substantially larger. The median REW and characteristic equivalent
width, $w^*$, are also both very much larger than the full
\Lya~sample.

Because the median REWs of sample~2 are significantly smaller than
that of the LWHED sample~6, we do not necessarily expect to detect
heavy element lines in many of the weaker \Lya~lines of sample~2,
given the FOS sensitivity, even if there were no significant
differences in metallicity.  One might expect therefore, that the
difference in rates of evolution between sample~6 {\it vs.} sample~2
simply reflects differing rates of evolution between low and high
equivalent width samples, and has very little to do with metallicity.

It would not be unexpected that in a subset of the Lyman--only lines
drawn from the high end of the equivalent width distribution of the
Lyman--only sample, we would obtain a rate of evolution which was
comparable to the LWHED samples, and we might then attribute
the lack of detected heavy element lines in such a sample to, for
example, the expected position of C~IV falling outside our wavelength
coverage.

It is not possible to find a sample of high REW Lyman--only systems
which exactly matches the equivalent width distribution of the LWHED
systems and still have a sample of reasonable size. However,
considering just lines in the Lyman--only sample with REW $>$
0.70~\AA~(samples~8 and~9 for the V and U samples, respectively) we
may compare the 0.25, 0.50 and 0.75 percentiles in the REW
distributions with the following results: For samples~6 and~8 (the V
samples for the LWHED and Lyman--only samples) the REWs for these
three percentiles are 0.48, 0.80 and 1.09 {\it vs.} 0.74, 0.83 and
0.99, respectively, while the corresponding comparisons for the
U~samples~7 and~9 are 0.63, 0.89 and 1.15 for the LWHED systems {\it
vs.} 0.74, 0.82 and 0.99 for the Lyman--only systems.

Comparison of the maximum likelihood values for $\gamma$ in sample~6
{\it vs.} sample~8 as well as sample~7 {\it vs.} sample~9 shows that
the LWHED systems seem to be evolving much more rapidly than the
Lyman--only systems, though the errors are large due to the small
number of lines. Utilizing the formal 1$\sigma$ errors given in Table
2, we find formal differences in the rate of evolution of the
Lyman--only and LWHED systems exist at about the 2.2$\sigma$ and
2.4$\sigma$ level for the V~and U~samples, respectively.

Recognizing that the apparent differences in rates of evolution
between the LWHED systems and the high end of the REW distribution for
the Lyman--only systems are not highly statistically significant it is
nevertheless of interest to consider whether selection effects might
artificially give rise to this apparent difference. Or, if the
difference is real, what the likely interpretation is.  Since a
detailed discussion of the properties of the heavy element systems in
the Key Project sample will be presented in a separate paper
(\cite{sargberg}), we will give only a brief and mostly qualitative
discussion.

There is one effect which might cause systems which we have considered
to be LWHED systems to actually be Lyman--only systems: Accidental
groupings of real \Lya~lines might lead to false identifications of
metal line systems satisfying all the requirements of the line
identification algorithm. As discussed in \S3, and in detail
in CAT2 and CAT3, when the candidate C~IV doublet is in the
\Lya\ forest, and when the \Lya\ line which would be present at the
same redshift as the candidate C~IV is beyond our spectral coverage,
then of order 50\% of these candidate C~IV systems may be spurious.
Substituting pairs of \Lya\ lines for the C~IV identifications thus
increases by a few percent the total number of \Lya\ lines in the
total \Lya\ sample (and in the Lyman-only sample), but, as discussed
in \S4, has only a very small effect on the maximum likelihood fit
parameters. Altering the identifications of such cases does not change
the composition of the LWHED sample, since, by definition, {\it both}
heavy element ions and \Lya\ must be present. However, the discussions
in CAT2 and CAT3 also demonstrate that when a candidate C~IV system is
found and when the \Lya~in the same system {\it is} accessible,
\ie~when it is a candidate LWHED system, then the number of false
identifications {\it per spectrum} is very small. Thus, given the
total number of LWHED systems in samples~6 and~7, it is unlikely that
misidentification of accidental spacings of a very small number of
\Lya~lines as heavy element lines significantly affects our results.
Conversely, one might also suppose that the lack of heavy element
detections in samples~8 and~9 simply reflects biases in the spectral
coverage, and that with much broader spectral coverage, but the {\it
same sensitivity}, nearly all the Lyman--only systems of samples~8
and~9 would have heavy element detections.

Granting that this could be the case however, we would still be faced
with the question of why the LWHED and these Lyman--only systems seem
to have different evolutionary rates, and whether the bias introduced
by the incomplete coverage is responsible for this difference.

The following argument suggests that biases in spectral coverage are
not likely the explanation for the effect we observe.  In the range of
rest wavelengths accessible with the gratings used in our sample, the
C~IV doublet is the most ubiquitous ion found, as is well known from
groundbased studies.  The other most prominent lines (or doublets)
are, in order of decreasing rest wavelength: the Si~IV doublet
($\lambda\lambda$1393,1402), C~II~$\lambda$1334, Si~II~$\lambda$1260,
the N~V doublet ($\lambda\lambda$1238,1242), Si~III~$\lambda$1206, the
O~VI doublet ($\lambda\lambda$1032,1037) and C~III~$\lambda$977.

If all of these ions were roughly equally likely to occur, then the
lack of spectral coverage would have little effect on the values of
$\gamma$ in samples~8 and~9. The Si~II~$\lambda$1260 line, for
example, representative of low ionization systems, is nearly always
accessible when \Lya~is itself.  The same is true for
Si~III~$\lambda$1206, typical of intermediate ionization systems.  The
lack of coverage which would lead to the inability to detect the high
ionization C~IV doublet for \Lya~redshifts above about 1.1 would be
roughly compensated for by the ability to detect O~VI for
\Lya~redshifts above about 0.65, for objects in which G190H data
exist, and above about 1.2 for those objects in which only G270H data
exist.

However, to the extent that C~IV is the most frequently occurring and
readily detectable heavy element ion, extending the spectral coverage
to wavelengths beyond $\lambda$3270 (corresponding to C~IV redshifts
above 1.1) would preferentially remove systems at high redshifts
($z>1.1$) in the putative Lyman--only systems and transfer them to the
LWHED systems. This would tend to steepen the LWHED evolutionary rate
and flatten the Lyman--only rate. To make a semi--quantitative
estimate of the magnitude of this latter effect, we define one
additional sample of Lyman--only lines (sample~10, Table~2) which has
precisely the same sensitivity limits and path length coverage as the
LWHED sample~7. From these two samples we calculate the fraction of
\Lya\ lines in which the C~IV ion is seen in those cases in which it
was accessible (about 12\%). For every \Lya\ line in sample~10 for
which the redshift is such that C~IV was {\it not} accessible we used
a random number generator and this fraction of 12\% to produce a set
of ``virtual line lists'' in which we imagined that the C~IV line was
sometimes detected in instances where it was not accessible in the
real line lists. The average values of $\gamma$ resulting from this
procedure for what we might call ``virtual samples~7 and~10'' (but not
shown in Table~2) are 1.95 and 0.16 respectively, (with uncertainties
in $\gamma$ about the same as the real values) compared to the actual
values of 1.55 and 0.25. The effect is in the anticipated direction---
increasing the spectral coverage would preferentially discover high
redshift C~IV lines, and move high redshift systems from the
Lyman-only category into the LWHED category, thus increasing the
difference in the two gammas.  We find that the difference in the
$\gamma$'s between these two virtual samples is significant at about
the 2.8$\sigma$ level.

One additional possibility which might give rise to the different
rates of evolution found for samples~6 {\it vs.}~8 or 7~{\it vs.}~9
would arise if the high ionization systems which cluster around the
emission redshifts, most of which appear to be intrinsic to the QSO,
have a significant high--velocity tail which extends well beyond
3000~km~s$^{-1}$. Removing such systems would preferentially remove
the highest redshift systems along each individual line of sight and
thus lower the $\gamma$ of the LWHED systems. Analysis of the actual
distribution in ``ejection velocity'' space shows that this is not
likely to be an important effect: In the 5000~km~s$^{-1}$ bin between
$-$2000 and 3000 km~s$^{-1}$ (the ``in-falling'' systems probably
represent the difference between the true systemic velocity and that
inferred from the broad emission lines) there are 19 LWHED systems. In
the next two 5000~km~s$^{-1}$ bins (from 3000 to 8000 and 8000 to
13,000 km~s$^{-1}$) there are 4 and 5 such systems respectively, and
the numbers remain at about this background level for the higher
velocity bins. Evidently, there is no very significant high velocity
tail which could likely give rise to the possibility just described.
 
While the foregoing is hardly a rigorous analysis, it seems unlikely
that effects primarily attributable to incomplete spectral coverage,
line misidentification, or a high velocity tail of intrinsic systems
can account for the LWHED rate of evolution being steeper than those
exhibited by the Lyman--only systems. Thus, if one is willing to
accept these results as statistically significant, what is the most
likely explanation for this?  It is of course possible that the LWHED
and high-REW Lyman--only systems really represent differences in metal
content.  Only by improved sensitivity, higher resolution, larger
samples and extended spectral coverage can this be confirmed. A
program which can and should be readily carried out is to obtain the
appropriate groundbased data in order to extend the data base for the
C~IV doublets which actually do accompany the \Lya~lines in our
samples~8, 9 and~10 above redshifts of 1.1.

We believe a more likely explanation for the difference in the
evolutionary rates between the LWHED and Lyman--only systems than
metallicity differences, is simply that there is a significant
difference in H~I column densities between the LWHED systems and the
Lyman--only systems of samples~8 and~9, despite our attempt to match
the two samples in terms of the distribution of equivalent widths. As
examination of the curves-of-growth in Figure~5 makes clear, for the
entire range of equivalent widths represented by nearly all the \Lya\
lines in samples~6--10, the equivalent widths give virtually no
reliable information at all about the H~I column density. This
insensitivity is worsened by blends which cause blended lower column
density systems to masquerade as very high column density single
systems.  The LWHED systems probably simply act as markers to pick out
those systems with H~I column densities large enough for the heavy
element ions to be detected at our FOS sensitivities. We would then
conclude that over the 2-3 orders of magnitude range in H~I column
densities represented by samples 6--10 there is a systematic increase
in the rate of evolution with increasing H~I column densities.
Support for this view comes from the rate of evolution of the Lyman
limit systems which occur fairly frequently among the stronger LWHED
systems having several different ion species (CAT3).  The Lyman limit
systems were found by \cite{steng95} to have values of $\gamma$ in the
range 1.0--1.5, depending upon the particular sample analyzed.

\subsection{Does the Rate of Evolution of the \Lya\ Lines Vary with \Lya\ REW?}
Despite the insensitivity of the REW to H~I column density for all but
our weakest lines, we can attempt to examine whether there is any
evidence within our sample for a dependence of the rate of evolution
upon rest equivalent width. In view of our conclusion above concerning
the reason for the difference in evolution rate between the
Lyman--only and LWHED systems, we ignore the distinction between the
Lyman--only and LWHED systems and divide up the 920 \Lya\ lines in
sample~2 into five REW bins, whose boundaries are chosen so that for
the five samples each bin contains the same number (184) of \Lya\
lines.  This leads to REW bin boundaries (in \AA) of 0.06, 0.22, 0.32,
0.45, 0.64 and $\infty$.  (There are no lines detected with REW less
than 0.06~\AA~in our sample.)  Except for the highest two REW bins,
the spread in rest equivalent width is not large enough to allow the
parameter $w^*$ to be well--determined for the individual bins. We
have therefore fixed the value of $w^*$ to the value found in sample~2
(0.273). The resulting values of $\gamma$ depend only weakly on the
actual value of $w^*$ selected, as long as this value does not fall
much below about 0.20. For the higher equivalent width bins, the
samples are virtually identical for the U and V samples, but for the
lower REW bins both the sample size and redshift range drop strongly
for the U samples (the U sample has no meaning for the lowest REW bin,
since there is virtually no coverage over which the minimum detectable
REW is everywhere $\leq$ 0.06~\AA.) We thus consider only V samples.
We estimate the errors in $\gamma$ due to the formal uncertainty in
the fit, together with the sensitivity to the assumed fixed value of
$w^*$, to amount to about $\pm$0.50. The resultant values of $\gamma$
{\it vs.} rest equivalent width are shown in Figure~7.  Figure~7 is
suggestive of an increase in the rate of evolution with increasing
equivalent width, in accordance with expectations if our preferred
explanation for the difference in the LWHED and Lyman--only systems is
correct, though the trend exhibited by the first 4 bins is not
continued for the highest REW bin. This result is qualitatively in
accord with the analysis of \cite{dobrbech97} as well as the analysis
of \cite{kim97} at high redshifts based upon high resolution data.

\section{Summary and Discussion}
\subsection{Summary of Results}
Using the homogeneous sample of HST--FOS spectra obtained during the
QSO Absorption Line Key Project, and the set of lines measured and
identified using objective and reproducible procedures, we have
analyzed a subset of 63 of these spectra to investigate the evolution
of the \Lya\ lines in the sample. The subset of these objects and
associated line lists are those objects excluding the strong broad
absorption line QSOs, QSOs whose spectral coverage did not cover the
region blueward of \Lya, QSOs whose line identifications are not yet
complete, and those QSOs or portions of QSO spectra that were only
observed using the low dispersion G160L grating, having a resolution
of $\approx$ 1100 km~s$^{-1}$.

Our main result is that over the redshift range from 0 to 1.5, the
rate of evolution, as parameterized by $\gamma$ in the expression
$(1+z)^{\gamma}$, is very much flatter than found from groundbased
data which involve redshifts higher than about 1.7 and for which
estimates in the literature have varied over a wide range of
values. In particular, the analysis by \cite{bechtold94} yielded a
value of $\gamma$ of 1.85 (see Figure~6). The fits to the groundbased
data and our data intersect at a redshift of about 1.5--1.6.

The values of $\gamma$ we find (\eg~0.16 for sample~2) are in general
smaller than what is expected for clouds whose product of (comoving)
number density and cross--section is constant, as might be expected
for, \eg~stable clouds in galaxy disks or halos. For the currently
popular value of $\Omega \approx 0.2$, under this assumption the value
of $\gamma$ varies only slightly, from 0.9 at zero redshift to about
0.85 at a redshift of 1.5. Our data thus implies that the overall
ensemble of \Lya\ absorbers evolve so that this product becomes
smaller with increasing redshift.

The fit to a power law in $(1+z)$ for the larger samples is quite
good, as is the fit for the distribution of rest equivalent widths by
an exponential.

We find two exceptions to the flat rate of evolution described above:

1) When we examine the \Lya\ lines from the single high redshift QSO,
UM~18, for which complete line identifications have been made, the
line density in the redshift bin between $z=1.5$ and $z=1.7$
apparently shows a strong upturn in the rate of evolution (though the
identifications are still somewhat uncertain and there are only 19
lines in this bin).

2) Samples consisting of \Lya\ lines which have ions of the heavy
elements identified at the same redshift as the \Lya\ lines, or which
are within 300 km~s$^{-1}$ of the redshift of systems containing such
ions (the LWHED samples~6 and~7) show a much faster rate of evolution
than the large total samples (\eg~sample 2) or the Lyman--only
samples (\eg~samples 4 and 8) even when a subset of the Lyman--only
systems crudely matching the rest equivalent width distribution of the
LWHED set are considered. Discussion of selection effects suggests
that line misidentification and incomplete spectral coverage is
probably not responsible for this difference.  In fact, simulations
suggest that the difference between the LWHED samples and the
Lyman--only samples increases when account is taken of the incomplete
spectral coverage of our data.

We find evidence that the rate of evolution increases with increasing
REW.

\subsection{Discussion}
As noted in the Introduction, there have been significant advances in
the detail with which simulations of the properties of the \Lya\
absorption lines can now be carried out. It is the case, though, that
high resolution simulations incorporating small (galaxy-sized) scales,
high column densities, and shocks, and requiring the incorporation of
both hydrodynamic and many--particle gravitational processes, have
largely been confined to redshifts above about 2.0, well above the
redshift regime of our data, and are just now being extended to
redshifts of 0.0.

There is a general consensus, though, that if one avoids the smallest
scales, and higher column densities, then the baryons trace rather
well the collisionless particles in the cold dark matter
scenarios. \cite{riedig98} (hereafter RPM) took advantage of this to
carry out simulations using a many--body code to obtain results to
$z=0.0$, allowing for shock heating in an approximate way. We
therefore limit comparison of our results to this paper, in which
comprehensive references to other recent simulations may be
found.\footnote{We have recently learned that a full hydrodynamic
simulation to zero redshift has been carried out by Dav\'e \etal\
1998.  This work also agrees with our result in finding a decrease in
the rate of evolution at redshifts of roughly 1.5, as well as agreeing
qualitatively with our result that the weaker lines evolve more slowly
than the stronger ones.}

By also modeling the evolution of the UV background radiation, which
obviously plays a major role in determining the HI column density,
their simulations yield the evolution of line density with redshift
(for lines with $\log N_{HI} \geq 10^{14}$) ---see their
Figure~5---which is quite similar to our Figure~6. (For a Doppler
parameter of 30 km~s$^{-1}$, this column density implies a rest
equivalent width of about 0.25~\AA---see our Figure~5). Since the
thermal history of the gas plays an important role in the simulations,
which is in turn affected by whether the gas has been shocked or not,
RPM distinguish two populations of absorbers (``$P_u$'' and
``$P_s$''), in which material in the first population has not been
shocked, and material in the second population has been.  (In reality,
there is no doubt a continuum of properties rather than a sharp
dichotomy between two such populations.)

The RPM simulations indicate that the unshocked material is found
primarily in under-dense regions, and in the more outlying regions of
condensations, while the shocked population is found in the vicinity
of regions undergoing condensations to clusters and individual
galaxies. RPM further suggest that it is among the shocked population
that metal enrichment will be found. Their simulations show that the
unshocked population evolves the more rapidly with increasing redshift
and thus dominates the absorbers at high redshifts, while it is the
shocked population that has the flat evolution and dominates at low
redshifts (see their Figure~6).  Finally, they suggest that the very
weakest lines evolve much more slowly than the stronger lines (their
Figure~7).  Their models are thus in qualitative agreement with our
results, but the column densities they associate with their shocked
population and which they predict to dominate at low redshifts are
smaller than we can detect with our FOS data.  While their association
of the metal--enriched lines with the slowly evolving shocked
population might seem in apparent contradiction with our result that
those \Lya\ systems with detectable heavy elements evolve more rapidly
than those without, the minimum degree of metal enrichment which these
authors envisage is well below our threshold for detection.

Regardless of whether the division of absorbers into ``shocked'' and
``unshocked'' populations is realistic, both the apparent rapid
evolution of the \Lya\ lines above a redshift of $\sim$ 1.5, and the
difference in rates of evolution of the Lyman--only and LWHED systems
suggest that rather different physical processes and/or environments
are involved, and these different processes and environments would be
expected to also affect the question of the association of \Lya\
absorbers with galaxies.

The dependence of the rate of evolution upon the \Lya\ line strength
suggested by our Figure~7 raises interesting questions about the
nature of the relationship between the \Lya\ absorbers and galaxies,
and in particular about the nature of this relationship for the weaker
lines. The relation between galaxies and low redshift \Lya\ absorbers
has been investigated by a large number of investigators over a number
of years. For example, \cite{bahcall79} suggested that moderate sized
galaxy halos could account for metal-containing absorption systems
while larger halos could account for the larger number of \Lya\
lines. Subsequent to the launch of HST, the relation between galaxies
and low redshift \Lya\ lines has been examined quantitatively by a
number of groups. In a recent extensive analyses, \cite{chen98}
examined the cross--correlation statistics between \Lya\ absorbers and
galaxies {\it for} \Lya\ {\it absorbers with equivalent widths greater
than 0.3}~\AA. They find that the cross section for producing
absorbers at a given equivalent width increases with the B-band
luminosity of the galaxy (but is not strongly dependent upon
morphological type) and that for a given luminosity there is a strong
anti--correlation between equivalent width and impact parameter. They
also find that {\it the absorption properties show little or no
correlation with redshift}. They conclude that a significant
fraction---at least 50\%--- of all such absorbers are physically
associated with outer gaseous envelopes of galaxies extending to
$\sim$160 kpc. They further suggest that very low luminosity
galaxies---whose properties and spatial density are not yet
well--determined---could account for the remaining fraction.

Our fits for the value of $\gamma$ of our large samples (\eg~sample
2) are significantly less than expected for objects whose product of
comoving number density times cross section is constant, and which, as
noted above, for the currently popular value of $\Omega \sim 0.2$,
implies a value of $\gamma$ of about 0.9. The assumption of a constant
product of comoving number density times cross section is, in fact,
implicit in the evaluation by \cite {chen98} of the expected \Lya\
line density---see their equation (29)--and is supported by the
absence of any significant correlation of galaxy--absorber properties
with redshift in their analysis.

Our conclusion in \S7.2 that the rate of evolution depends upon line
strength, and in \S5 that the product of number density times
effective cross section decreases with increasing redshift for the
sample as a whole, leads to at least one of the following three
conclusions:

(1) The association between \Lya\ absorbers and galaxies {\it at
the lower end of
our equivalent width distribution} is not as
tight as implied
by the \cite{chen98} relation. Inspection of contours of constant
H~I column density which have been published for higher redshifts
suggests this is likely to be the case, and for the weakest low
redshift \Lya\ lines the notion of well-defined ``galaxy halos'' is
probably not very meaningful.  

(2) Over the redshift and equivalent
width ranges of our sample, there is a more complex and
redshift-dependent relation between line strength, galaxy luminosity
and impact parameter than that found by \cite{chen98} over the more 
limited range of line strength and redshift covered in their analysis.

(3) The types of galaxies associated with a significant number of the
\Lya~lines evolve in a very different way from ``ordinary'' galaxies.
In particular, the rate of evolution we deduce for the weaker lines in
our sample is different from the population of blue dwarfs, (which
\cite{chen98} suggest may account for a significant fraction of the
\Lya~lines), whose rate of evolution is {\it more} rapid than that
of typical luminous galaxies.

Conclusion 1) would be consistent with the
failure to find a strong galaxy--\Lya\ absorber association among the
very low column density absorbers (\cf\ \cite{morris93}).
Only more detailed studies of the statistics of galaxy-\Lya\ association
extending to weaker \Lya\ lines, higher redshifts, and fainter galaxies can definitively settle these issues.

We suggest that the LWHED systems involve moderate H~I column density
clouds, higher than those associated with the weaker Lyman--only
systems, but generally not as high as the low--ionization and Lyman
limit systems, which comprise a relatively small fraction of the LWHED
sample. These latter systems may arise in clouds pressure-confined in
thermalized halos, as initially suggested by \cite{bahcall69}; investigated observationally by, e.g. \cite{berg91} and \cite{steid94}; and
discussed in detail theoretically by \cite{mo96}. These absorbers may
have column densities which are too high and scales which are too
small and too closely associated with galaxies to be well--reproduced
in the RPM simulations.

The foregoing, together with the results of \S7.2, imply that at low
redshifts, high H~I column density clouds evolve more rapidly than
those of low H~I column density, just as they do at higher redshifts
(\cite{kim97}). However, regardless of what  combination of the three
scenarios enumerated above proves most nearly correct, it still leaves
unanswered the question of {\it why} the line density evolves as it
does as a function of redshift and line strength. The variation of the
ionizing background radiation with redshift obviously plays an
important role, but cannot be the sole factor, since such variations
would cause the H~I column densities to scale in the same way. However,
local evolving ionizing sources could bring about differing rates of
evolution by affecting more strongly the nearer and higher H~I column
density clouds.

Our conclusions and speculations need to be, and can be, investigated
more carefully using high resolution and high sensitivity data.  In
particular, at the redshifts dealt with in this paper, a data set
taken with, e.g. STIS, at a resolution of $\sim15$ km~s$^{-1}$ would
avoid almost entirely ambiguities associated with line blending.  A
rigorous assessment of the effects of line blending at the FOS
resolution would require detailed simulations of spectra which cover
the full range of line strengths and density of lines encountered in
our data set, and which utilize the relatively small amount of
existing high resolution HST data. Such simulations would then need to
be analyzed using exactly the same procedures we have used for the
actual data. A few such simulations and analyses have been carried out
(Paper~II), but not at sufficiently high line densities. Lacking this,
we give only the following qualitative and semi-quantitative comments:

At low redshifts (\eg~below 0.5) the observed line densities in our
spectra are so low that blending of lines from the mean background is
not a significant problem, nor is the effect of line crowding in
affecting the placing of the continuum. For example, in the spectrum
of PKS~0405$-$12, the mean separation between the 4.5$\sigma$ lines in
our sample is about 15 times the spectral resolution, so blending by
random superposition of lines rarely occurs. It could nevertheless be
the case that even isolated features break up into multiple components
if there is substantial velocity structure on scales less than our
resolution. Resolving such structure would of course affect the
derived equivalent width distribution, but unless the amplitude of
such structure varied with redshift it would not affect the
discussions of the rate of evolution. In any case, the rather sparse
amount of high resolution data on the very low redshift \Lya\ lines
suggests that this is not likely to be an important effect.

At the upper end of our redshift range ($\sim$1.5), the ratio of the
mean line separation to the spectral resolution has decreased to about
five, for the reasons discussed in \S4. Thus, quite aside from
clustering, we should expect that of order 20\% of our 4.5$\sigma$
lines would be affected by blending. Depending upon the detailed
structure of such blends, the deblending algorithm may or may not
break up such blends into the correct number of components, and the
assignment of the individual equivalent widths to the components is
dependent upon the detailed structure of the blend.  The effect of
blending upon the rate of evolution of the \Lya\ lines is subtle and
complex. Naively, one would expect that blending would decrease the
number of lines compared to the number measured with higher resolution
but the same limiting equivalent width.  However, resolving the blends
in such a situation also has the opposite effect, since the decreased
equivalent width of resolved blends would cause many of the lines now
in our sample to move below the 4.5$\sigma$ level. In addition to the
affects of line blending {\it per se} crowded spectra make the
placement of the continuum increasingly subjective.

Because of these uncertainties, in our comparison of the HST results with 
ground-based results, we have chosen to place more emphasis on the \cite{lu91}
and \cite{bechtold94} results, rather than higher resolution ground-based
data (\eg~\cite{kim97}) since the former two studies were carried out with
spectral resolution and line densities more nearly comparable to ours.

Finally, the foregoing discussion warrants strongly re-emphasizing the
remarks made in \S3.1: In our view, the value of our analysis and data
set will be most fully realized via comparison between the simulated
spectra of detailed hydrodynamic numerical simulations just now
reaching zero redshift. Such simulated spectra can be modeled to have
exactly the resolution and sensitivity characteristics of our data, and
{\it by analyzing them in precisely the same way that our FOS data set has
been analyzed, most of the problems above will be circumvented.}

In addition to largely avoiding problems caused by line crowding, if
the nature of the \Lya\ absorbers really can be described roughly in
terms of two components---or perhaps three as suggested above---but
more realistically in terms of a continuum of environmental conditions,
then a high resolution, high sensitivity data set would allow
exploration of other properties in addition to metallicity and rate of
evolution which would likely reflect these differing conditions. These
would include the correlation of the \Lya\ absorbers with galaxies or,
more generally, the amount of local over-- or under--density, and the
clustering among the lines themselves, as well as their Doppler and
equivalent width distributions. In CAT2 we have already presented
evidence indicating that some \Lya\ lines tend to cluster around strong
heavy element--containing systems.

The redshift regime around 1.4--1.8 appears to be an extremely interesting
transition region insofar as the \Lya\ absorbers are concerned, and
efforts to study correlations in these various properties in this
redshift regime should be carried out.  With the new generation of
large groundbased telescopes, and careful attention to the design of
the spectrograph and detectors, \Lya\ lines down to redshifts as low
as $\sim$ 1.55 can probably be studied more effectively than from
STIS, but for redshifts below that we will have to rely on STIS or the
Cosmic Origins Spectrograph. 

RJW, BTJ, and LL thank R.F. Carswell, M. Rauch, L. Storrie-Lombardi
and D. Weinberg, and RJW thanks L. Cowie and E. Hu, for helpful
discussions and comments on early drafts of the manuscript.

\newpage

\begin{appendix}
\section*{Appendix}

We use maximum--likelihood estimation to obtain values of the
parameters characterizing the evolution of the \Lya\ lines. A detailed
description of this formalism as applied to the evolution of the \Lya\
forest was given by \cite{murdoch86}. Our formalism follows in general
that presented by \cite{murdoch86} but differs in a few important
details, so we summarize here the method used in the present paper.

We parameterize the distribution function for the
\Lya\ lines by:

$${\partial^2N \over \partial z\partial w}=A\times G(z)\times H(w) \eqno(A-1) $$

\noindent
where A is a normalization factor and we assume a simple
power law distribution for G(z), and an exponential
distribution for H:

$$ G(z)=(1+z)^{\gamma}   \eqno(A-2) $$

\noindent
and

$$ H(z)= e^{-w/w^*} \eqno(A-3) $$

\noindent
where $z$ is the redshift and $w$ is the {\it rest} equivalent width (REW).
In \S5 we discuss whether these simple assumptions are
an adequate description of the distribution function. The extension of
the formalism described here to more complex expressions for $G(z)$ and
$H(w)$ is straightforward but for the present data set they seem adequate.

In order to explicitly obtain error estimates for the normalization
constant, A, we modify the \cite{murdoch86} formulation of the maximum
likelihood estimation so that the normalization factor A,~\ie\ the
line density at some fixed redshift, (generally chosen to be zero) and
integrated over some range of REWs, appears explicitly as a parameter
to be determined along with, \eg~$\gamma$ and $w^*$.

We imagine redshift--REW space to be divided up into a large number of
cells, small enough that we may always neglect the probability of two or
more lines being found in any one cell. If $\Delta \, n$ is the
expected number of lines to be found in a cell at redshift $z$ and REW
$w$, then the probability of the cell having no lines in it is
$P_{\rm empty}=e^{-\Delta n}$ while the probability of a cell being
``full'' (\ie~having one line in it) is $ P_{\rm full}=e^{-\Delta n}\times
\Delta n$.

Then the probability of the ensemble of the observations yielding a set
of lines filling the particular cells  $(z_1,w_1) \, , (z_2,w_2) \, ...$ is

$$  P = \prod_{\rm empty-cells}e^{-\Delta n}\times \prod_{\rm full-cells}
(\Delta n\times e^{-\Delta n}) \eqno(A-4) $$

The quantity to be minimized is the ``entropy'', $ S = -\ln P $ which can
be written
$$ S = -\ln P = \sum_{\rm all-cells}\Delta n - \sum_{\rm full-cells}
\ln(\Delta n) \eqno(A-5) $$

The expected number of lines, $\Delta \, n$, in any given cell whose center
is at $z, w$ is given by

$$   \Delta n=A\times G(z)\times H(w)\times\Delta z\times \Delta w \eqno(A-6) $$

We can replace the first sum in equation A-5 as an integral over all the 
accessible regions of redshift and REW space. Retaining only the
terms which involve the unknown parameters, one obtains for the function
to be minimized

$$ S =A\sum_{k=1}^{N_{QSO}}\int_{z_{low,k}}^{z_{upp,k}}G(z,\gamma, ...)
\Lambda(z, w^*, ...)b(z) \ dz - \sum_{i=1}^{N_{\rm lines}}
\ln(A\times G(z_i)\times
H(w_i)) \eqno(A-7) $$

\noindent
where $z_{low,k}$ and $z_{upp,k}$ are the
usable redshift limits for the $k-th$ QSO (see Table~1),
the first sum is over all the QSOs in the sample, and the
second sum is over all the lines in the sample. The quantity $\Lambda$
is defined as

$$ \Lambda =\int_{w_{\rm min}(z)}^{w_2}H(w) \ dw \eqno(A-8)  $$

\noindent
provided $w_{\rm lim} \le w_2$, and is zero otherwise, and where
$w_{\rm min}$ is the greater of $[w_1 \, , \, w_{\rm lim}]$. The two values
of the REW, $w_1$ and $w_2$, define any particular sample of lines
satisfying $ 0 \le \, w_1 \, \le \, w \, \le w_2 \, \le \, \infty$.

The function $w_{\rm lim}(z)$ is the REW at the wavelength corresponding
to \Lya~at redshift $z$ that yields a 4.5$\sigma$ detection for the
QSO in question. These functions are taken from the Catalog.

The quantity $b(z)$ is the ``blocking factor'' defined in CAT1. Note
that the normalization factor $A$ appears explicitly as an unknown
factor with an associated uncertainty, in contrast to the formalism in
Murdoch \etal~ The normalization constant $A$, along with $\gamma$,
and $w^*$ are found by finding their values for which $S$ is
minimized. In the case of the normalization constant, this can be done
explicitly and setting $\partial S/\partial A \, = 0$, and solving for
$A$ leads to

$$ A={N \over{\sum_{k=1}^{N_{QSO}}\int_{z_{low,k}}^{z_{upp,k}}
G(z,\gamma, ...)\Lambda(z,w^* ...)b(z) \ dz} }\equiv {N \over {Q(\gamma,w^*)}} \eqno(A-9) $$

\noindent
where $Q$, the effective path length, is defined by equation
A-9. Equation A-9 is an obvious result, but this formulation leads to
formal estimates in the uncertainty in A and the coupling between its
uncertainty and the uncertainty in the other parameters, as described
below. Note that the effective path length, $Q$, takes into account
the variation in signal--to-noise as a function of wavelength for each
individual spectrum, and couples the equations determining $\gamma$
and $w^*$ in a complex way. This expression for the path length also
depends explicitly, of course, on the assumption of the distribution
of rest equivalent widths being adequately described by an
exponential, leading to variable REW detection limit (``V'') samples.
To avoid this assumption we can restrict the spectral regions which
contribute to the path length and to the lines in the sample to those
regions having a detection limit which is everywhere more sensitive
than some specified value, and consider only lines above this value
(uniform detection limit, or ``U'' samples). As described in the text,
the results for the two cases are generally quite similar for large
samples.

The expression for A in equation A-9  can then be substituted into
equation A-7 and values for $\gamma$ and $w^*$ found. 

We have found it most convenient to solve for these remaining
parameters by using the simplex minimization routine AMOEBA as given
in \cite{numrec}.
\medskip
\centerline{\bf Error Estimation}
In this section we describe the method we have used for estimating
the uncertainties in the fitting parameters, but also describe two other
slightly different approaches, and compare the three methods to get some
indication of the reliability of the error estimates.

In general, if $P(A,w^*,\gamma)$ is the (normalized) probability distribution for the 3 fitting parameters and (for example) 
$$P^{\prime}(\gamma) =
\int_0^{\infty} P(A,w^*,\gamma) \ dA \ dw^* \eqno(A-10) $$
is the marginal distribution of $\gamma$, if we have no {\it a priori} knowledge of the other parameters, then lower and upper confidence limits
($\gamma_{lo}$ and $\gamma_{up}$ for $\gamma$) are given by the expressions
$$ C_{lo} \, = \, \int_{-\infty}^{\gamma_{lo}} P^{\prime}(\gamma) \ d\gamma   \eqno(A-11a)$$
and
$$ C_{up} \, = \, \int_{-\infty}^{\gamma_{up}}  P^{\prime}(\gamma) \ d\gamma  \eqno(A-11b) $$.

\noindent
whereas the dispersion in the value of $\gamma$ is, by definition,

$$ \sigma^2(\gamma) = \int_{-\infty}^{+\infty}
P^{\prime}(\gamma) \times (\gamma - \gamma_{\rm avg})^2 \ d\gamma \eqno(A-12)$$.

For a large number of lines in the sample, and for a large range in
redshift we expect the ``entropy'' to have a sharp minimum around the
most probable values and to rise steeply around this minimum, so that
expanding the entropy function around the most probable value and
keeping only 2nd order terms should be a good approximation. In this
case the probability function is Gaussian with the surfaces of constant
probability being ellipsoids. The 1$\sigma$ confidence limits, with $C_{\rm lo}
=0.1587$  and $C_{\rm up} = 0.8413$, then yield the same estimates as the
expression for the dispersion in equation A--12.

In general, the principal axes of the error ellipsoid do not
coincide with the A, $\gamma$ and $w^*$ axes (\ie\ there are non--zero
mixed 2nd derivatives in the expansion of the entropy function) and
the integral in A--12 is most conveniently evaluated by transforming
to the principle axes and then carrying out the appropriate
integrals. This method of estimating the errors in the maximum
likelihood function and using the Gaussian approximation
is the one used for the uncertainties quoted in
Table~2 and we refer to this method as ``method 1''.

For relatively small samples, the Gaussian approximation becomes
poorer, and the surfaces of constant probabilities typically take on a
``banana shape'' (\cf~\cite{storrie96}). As in
Storrie-Lombardi \etal, estimates for the uncertainty in the values of
the fitting parameters have
sometimes been made by locating the entropy contour for
which $S - S_{min} \, = \, 0.5$ and taking
\eg~$\sigma_{\gamma}$, to be the extremes of $\gamma$ defined by this
contour. We denote this method by ``method 2''.

When the Gaussian approximation is not used, the marginal distribution
function is not symmetric about the most probable value and the
average of the marginal distribution function is not the most probable
value. An alternative to method 2 is to use equations A-11a and A-11b
to evaluate the confidence limits. This method is somewhat preferable
to method 2 since it uses the distribution of S over the entire
parameter space. Rather than evaluating the integrals A-11a and A-11b
numerically over the probability function based upon a single
realization---\eg~the fixed (observed) number of lines along each line
of sight, and the fixed observed distribution of those lines in
redshift-equivalent width space---we regard the values determined from
the maximum likelihood estimation as representing some true underlying
distribution.  We then use Monte Carlo methods to: (i) sample the
number of lines along any sightline from a Poisson distribution whose
expectation value is that appropriate for the sensitivity and spectral
coverage of that sightline (ii) select the equivalent width and
redshift for the lines along this sightline from the underlying
$\gamma$ and $w^*$ distributions (iii) for each realization, use the
maximum likelihood code to give values of $A$, $\gamma$ and $w^*$. The
ensemble of realizations then populates empirically the density of
points in $A$, $\gamma$, $w^*$ space. One may then simply sort the
resultant values of, \eg~$\gamma$ from all the trials, and by
inspection determine the values corresponding to the two confidence
limits. This procedure probably comes closest to giving a true
estimate of the uncertainties in the parameter estimates and also has
the advantage that it serves as a check on the code. We refer to this
procedure as ``method 3''.

In practice, for the full data set discussed in this paper, and for the
determination of all three parameters, method 3 requires a prohibitive
amount of computation, and even method 2 is rather cumbersome.
In order to compare these three methods therefore, we have considered a
very simple case involving a single line of sight, with a uniform
detection limit, so that only the A and $\gamma$ variables are coupled.

Specifically, imagine a line of sight spanning the redshift range
$z=0.0$ to $z=1.5$ with an underlying value of $\gamma$ = 1.0, and
consider three cases for which the expected number of lines is 10, 100
and 1000.  For methods 1 and 2 we take the sum of $\ln(1+z_i)$ (see
equations A--2, A--6 and A--7) over the lines in the sample to be the
mean expected for $\gamma$ = 1.0. The results for these three methods
for estimating the uncertainty in $\gamma$ are summarized in
Table~A-1. For method 1, the values of $\sigma^+$ and $\sigma^-$ are
equal and the uncertainty goes exactly as $1/\sqrt{N}$, but for
methods 2 and 3 the upper and lower confidence limits are slightly
different and the scaling of the confidence limits as $1/\sqrt{N}$ is
only approximate. Inspection of Table~A-1 shows, however, that for the
sample sizes dealt with in this paper, the method we have used should
give results which are accurate to within a few percent.

\centerline{\bf The ``True'' Distribution Functions in Redshift and Rest
Equivalent Width} 

If the minimum detectable equivalent width at every redshift were
uniform, and if there were an equal number of sightlines at every
redshift, then the raw distribution functions would represent a fair
sample of the true distributions of lines as a function of equivalent
width and redshift. In general, neither of these conditions is
satisfied. As mentioned in the text, we can calculate cumulative
distribution functions in terms of ``effective redshift path lengths''
(as in \cite{storrie96}) and ``effective equivalent width
path lengths'', in which case every detected line increments the
empirical distribution function by an equal amount. In practice,
however, these path lengths are highly non--linear functions of
redshift and equivalent width. We have therefore elected to weight
each line to compensate for the incompleteness in the redshift and
equivalent width coverage.

We define a ``corrected cumulative distribution function''
for the redshift distribution (for the ``V'' sample) by
$$ C(z) = 0 \ \ \ \ \ \ \ \ \ \ \ \ \ \ z< z_1 \eqno(A-13a)$$
$$ C(z) = 1 \ \ \ \ \ \ \ \ \ \ \ \ \ \ z> z_N  \eqno(A-13b)$$
$$ C(z) = \sum_{j=1}^{k}1/\psi(z_j)
\ \ \ \ \ \ \ \  z_k < z < z_{k+1} \eqno(A-13c)$$

\noindent
where $z_j$ is the $j^{\rm th}$ redshift and there are $N$ lines in the sample,
and where

$$ \psi(z_j) = \sum_{m}^{N_Q} exp[-w_{\rm det,m}(z_j)] \eqno(A-14)$$

In equation A-14 the sum is over all the QSOs whose spectra include the 
wavelength corresponding to a \Lya\ line of redshift $z$, and $w_{\rm det,m}$
is the $4.5\sigma$ detection REW at that wavelength for the $m-th$ QSO.
(The normalization
to 1.0 is performed after the initial sum is calculated.)

For  the corrected equivalent width distribution the analogous expressions
are given by

$$ C(w) = 0 \ \ \ \ \ \ \ \ \ w < w_1  \eqno(A-15a)$$
$$ C(w) = 1 \ \ \ \ \ \ \ \ \  w> w_N  \eqno(A-15b)$$
$$ C(w) = \sum_{j=1}^{k}1/\Lambda(w_j)
\ \ \ \ \ \ \ \ \ \ w_k < w < w_{k+1} \eqno(A-15c)$$

\noindent
where $\Lambda(w_j)$ is given by

$$ \Lambda(w_j) = \sum_{m}^{N_Q}
\int_{\lambda_{min,m}}^{\lambda_{max,m}} \phi(w_{det,m}(\lambda), w_j)
\times (1+z)^{\gamma} \ d\lambda \eqno(A-16) $$

\noindent
and where $\lambda_{min,m}$ and $\lambda_{max,m}$ are the usable
minimum and maximum wavelengths for each QSO (corresponding to the
\Lya\ redshifts of Table~1) and $ \phi(w_{det,m}(\lambda), w_j) $ is 1
for $w_j > w_{det,m}$ and 0 otherwise.

These formulations give a reasonable representation of the ``true''
distribution functions, but give large weights to lines found in
sparsely sampled portions of redshift or equivalent width space
(Figure~2). This can be seen, for example in the jump in the
cumulative redshift distribution function (Figure~3) in the region
around redshifts from about 0.31 to 0.36, where the coverage from the
G130H grating stops and the region from 1600~to 1650~\AA~in the
G190H grating has very low sensitivity.
\end{appendix}

\newpage
\begin{figure}
\centerline{\psfig{figure=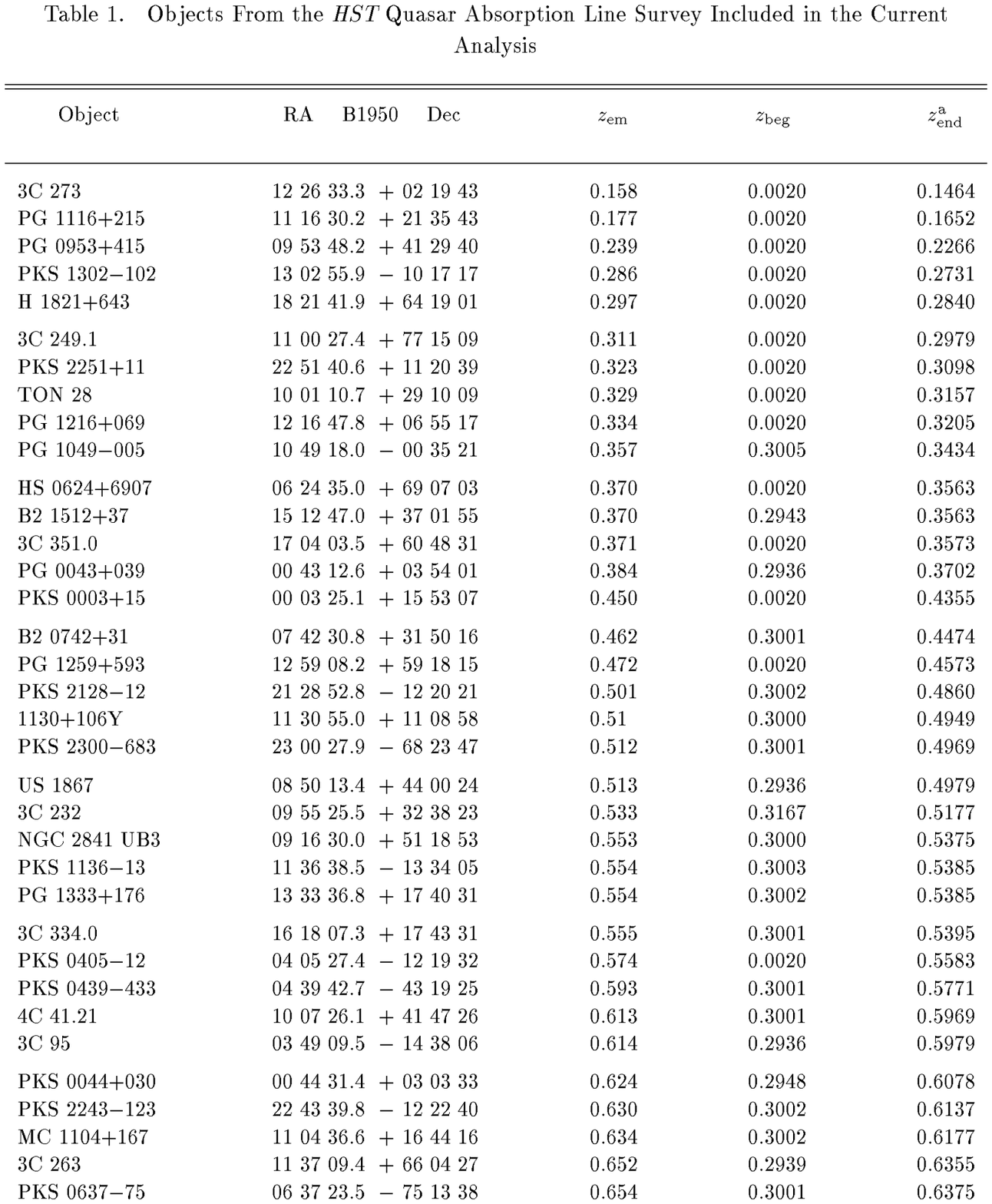}}
\end{figure}

\begin{figure}
\centerline{\psfig{figure=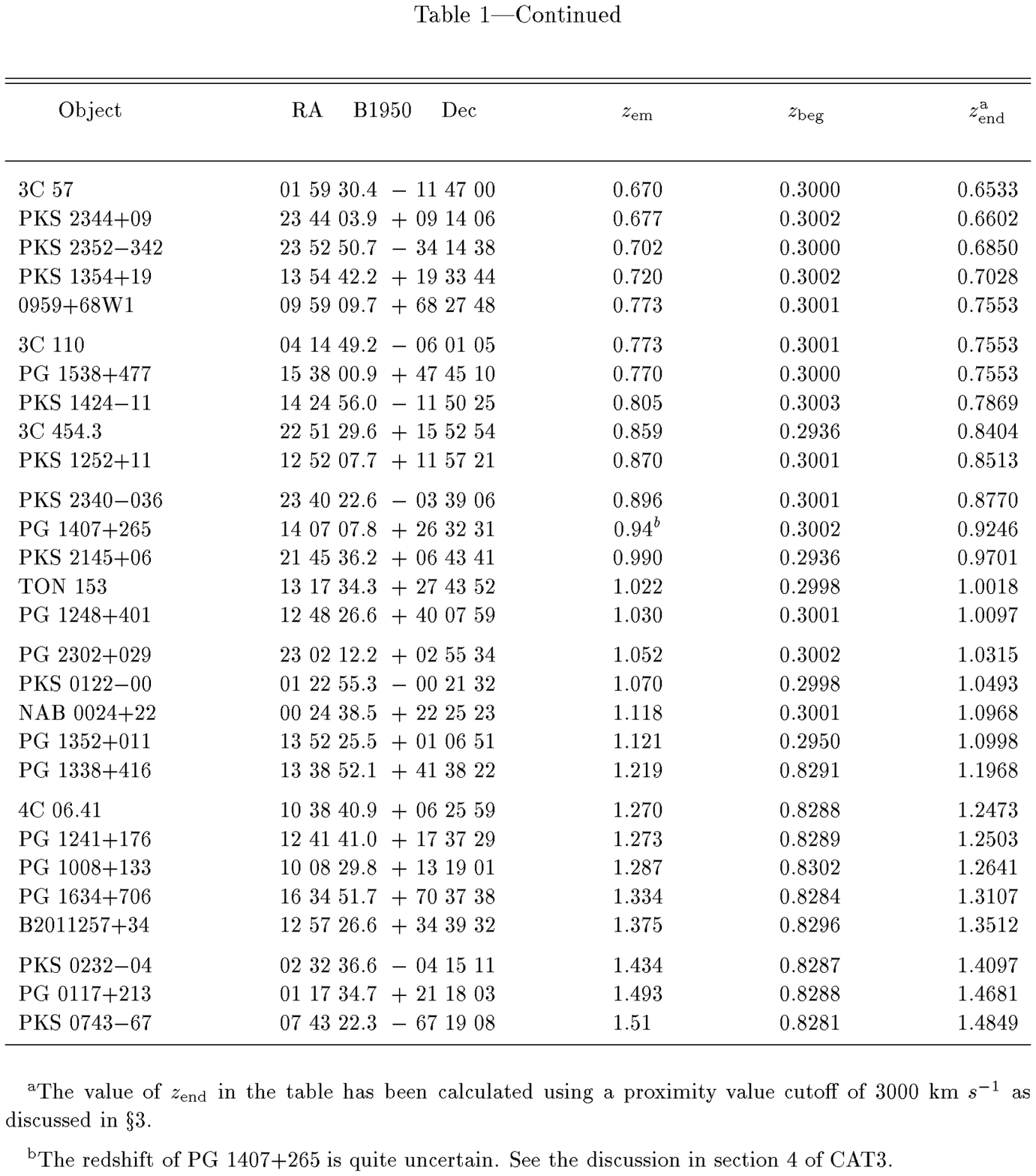}}
\end{figure}

\begin{figure}
\centerline{\psfig{figure=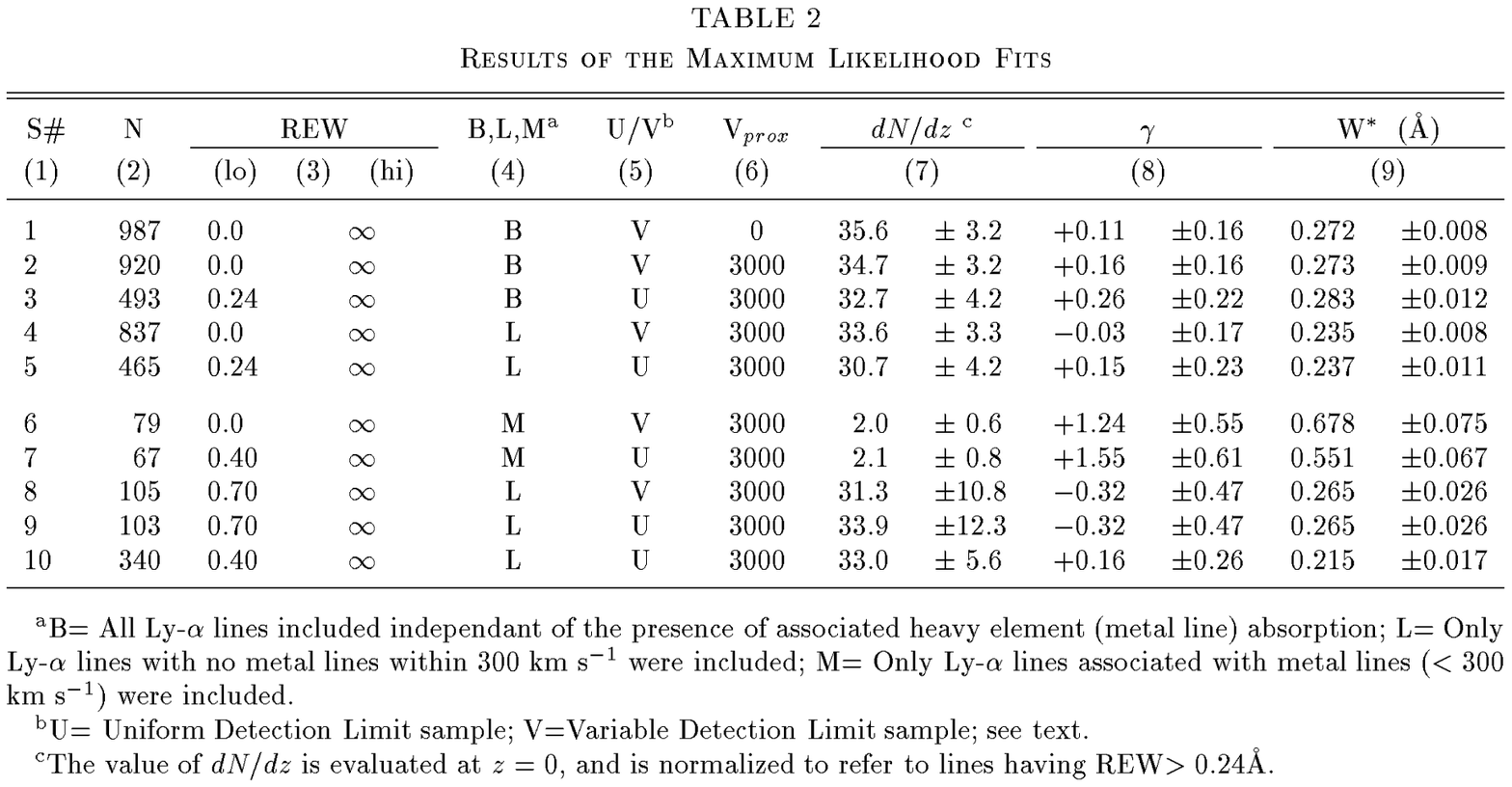,width=6.7in}}
\end{figure}

\begin{figure}[b]
\centerline{\psfig{figure=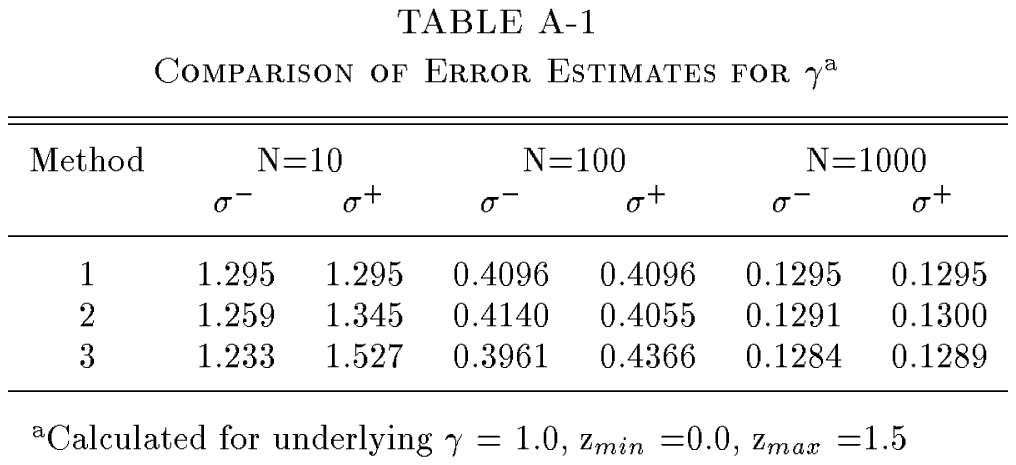}}
\end{figure}

\begin{figure}[t]
\centerline{\psfig{figure=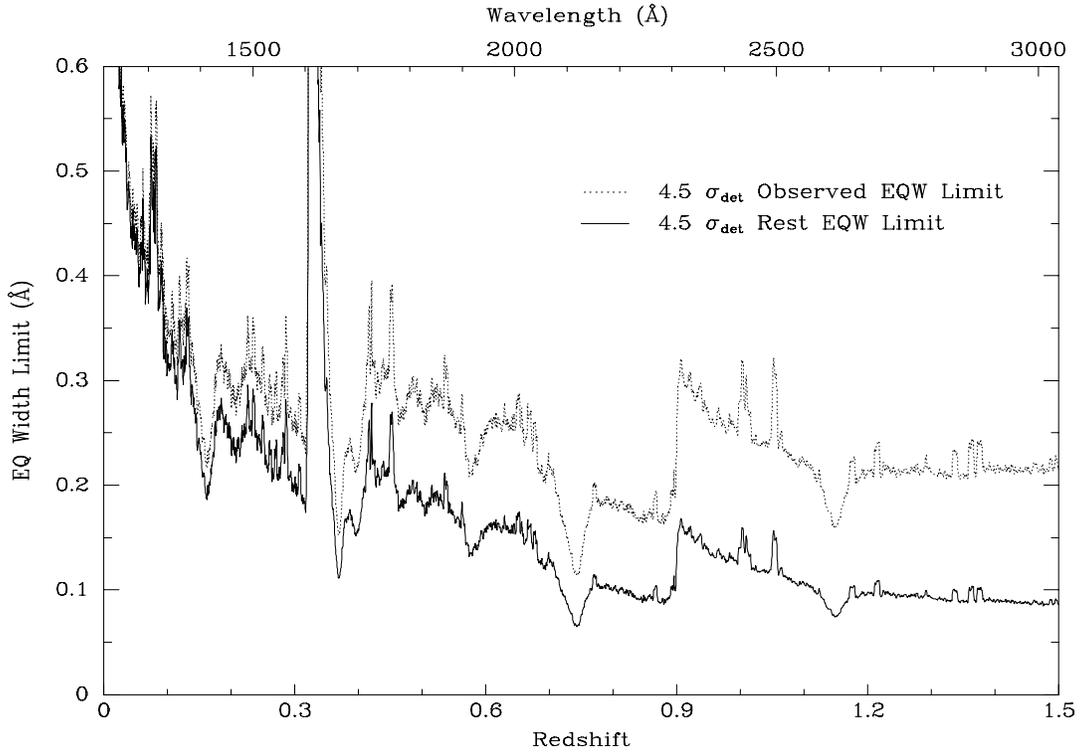,width=5.5in}}
\caption[]{\footnotesize Typical 4.5$\sigma$ 
equivalent width detection limits for
the three gratings (G130H, G190H, G270H) used in this study. The
usable portion of G130H covers the wavelength range from about
1220~to 1600~\AA, while the usable ranges of the G190H and G270H
gratings are about 1650~to 2300~\AA, and 2300~to
3270~\AA. The dotted line is the {\it observed} equivalent width limit
and the solid line is the {\it rest} equivalent width limit.
}
\end{figure}

\begin{figure}[p]
\centerline{\psfig{figure=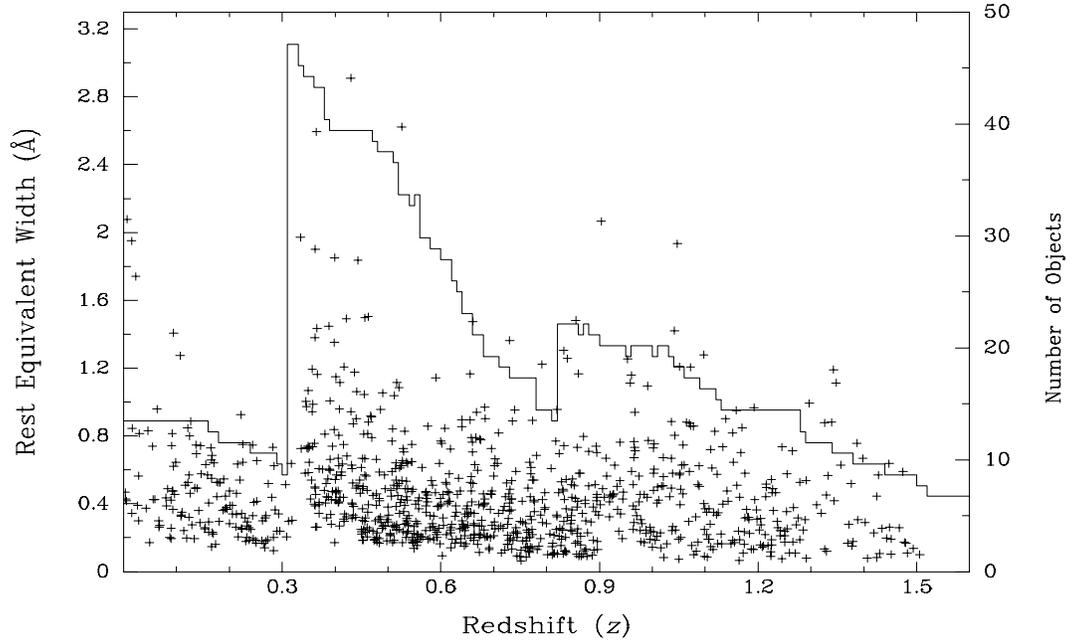,width=5.5in}}
\caption[]{\footnotesize The observed distribution 
of the 987 \Lya\ lines of sample~1
as a function of redshift and rest equivalent width. The histogram
gives the number of lines of sight in the entire Catalogue as a
function of redshift.
}
\end{figure}
\begin{figure}
\centerline{\psfig{figure=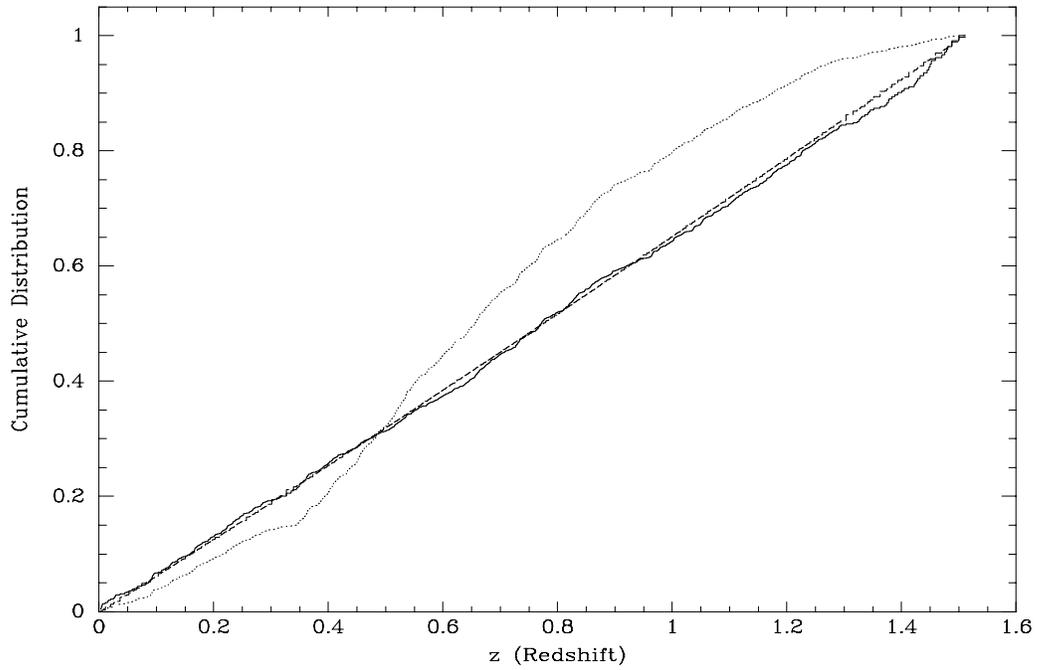,width=6in}}
\caption[]{\footnotesize The cumulative distribution function for the lines of
sample~1 {\it vs.}~redshift. The solid line is the cumulative
distribution corrected for incompleteness, as described in the
Appendix. The dashed line is the maximum likelihood fit to a power law
in $(1+z)$, while the dotted line represents the raw data.
}
\end{figure}

\begin{figure}[p]
\centerline{\psfig{figure=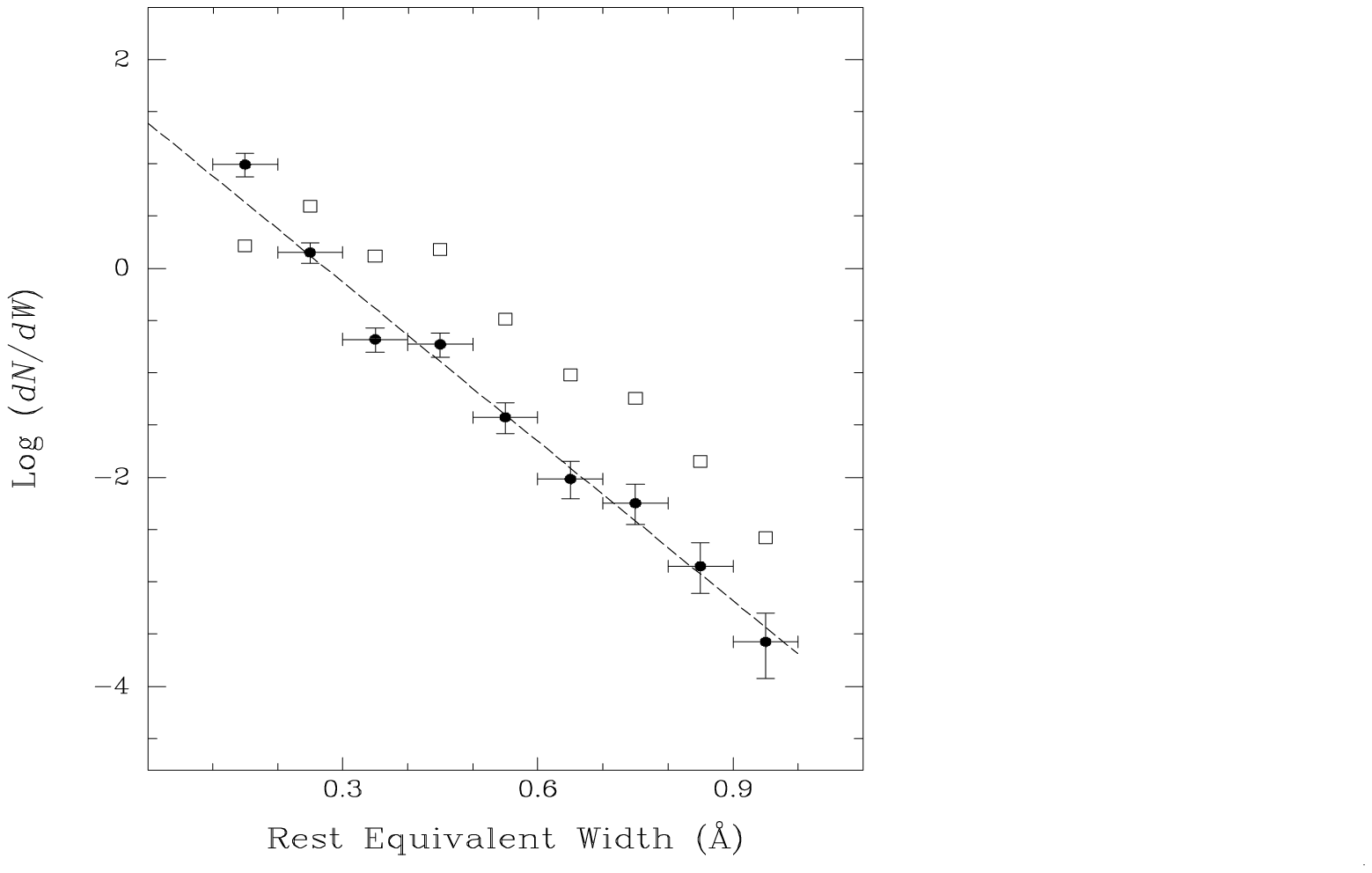,width=4in}}
\caption[]{\footnotesize The (differential) binned 
distribution ($\log_{10} dN/dw$)
of lines {\it vs.} rest equivalent width for sample~1, for nine
intervals of width 0.1~\AA~between 0.1~and 1.0~\AA. Open symbols
represent the raw data, while the closed symbols are the data
corrected for incompleteness, as described in the Appendix. The error
bars for the corrected distribution reflect the root~$N$
statistics. The dashed line is the maximum likelihood fit to an
exponential distribution. The scale for the ordinate indicates only
relative, not absolute, numbers, and the raw data have been shifted
relative to the corrected data for clarity.
}
\end{figure}
\begin{figure}
\vglue.3in
\centerline{\psfig{figure=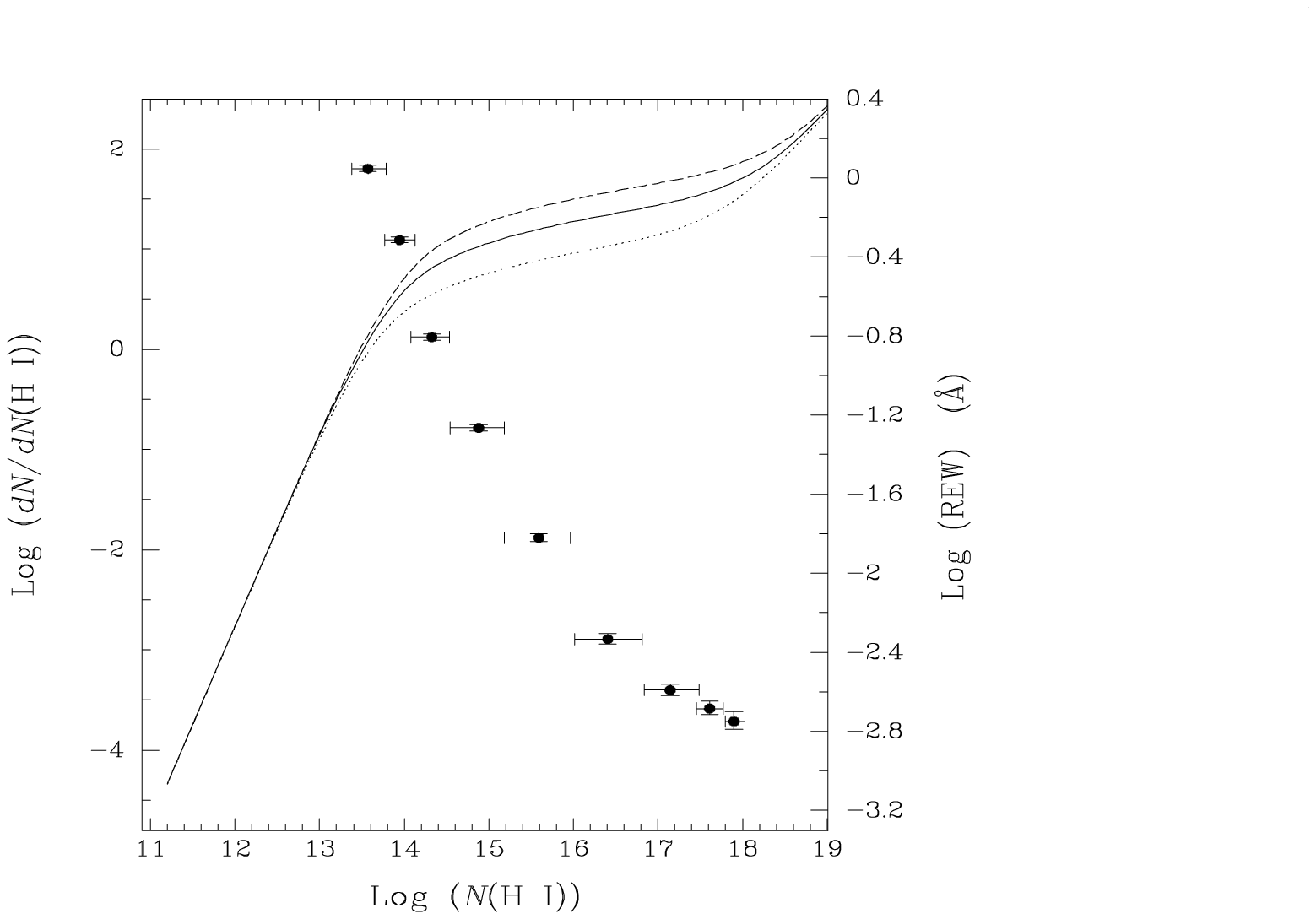,width=4.3in}}
\vglue-.1in
\caption[]{\footnotesize Three \Lya\ curves of growth for values of the Doppler
parameter of 40 (dashed line), 30 (solid line) and 20 (dotted line) km
s$^{-1}$, (right scale) and an approximate conversion of the plot in
Figure~4 from equivalent width to H~I column density. The solid
symbols represent the transformation of the binned corrected
distribution in rest equivalent width of Figure~4 to $\log N_{HI}$
column density, assuming a Doppler parameter of 30 km~s$^{-1}$.  The
lefthand scale for the ordinate indicates only relative, not absolute,
numbers. The extreme sensitivity of the inferred H~I column density
for given equivalent width to the Doppler parameter means that the
last several points at the high column density of the Figure are not
to be taken literally.  The approximate slope log $(dN/\,dN(\hbox{H~I}))$
for the weakest lines is about -1.3. See the text for further
discussion.
}
\end{figure}

\begin{figure}[t]
\centerline{\psfig{figure=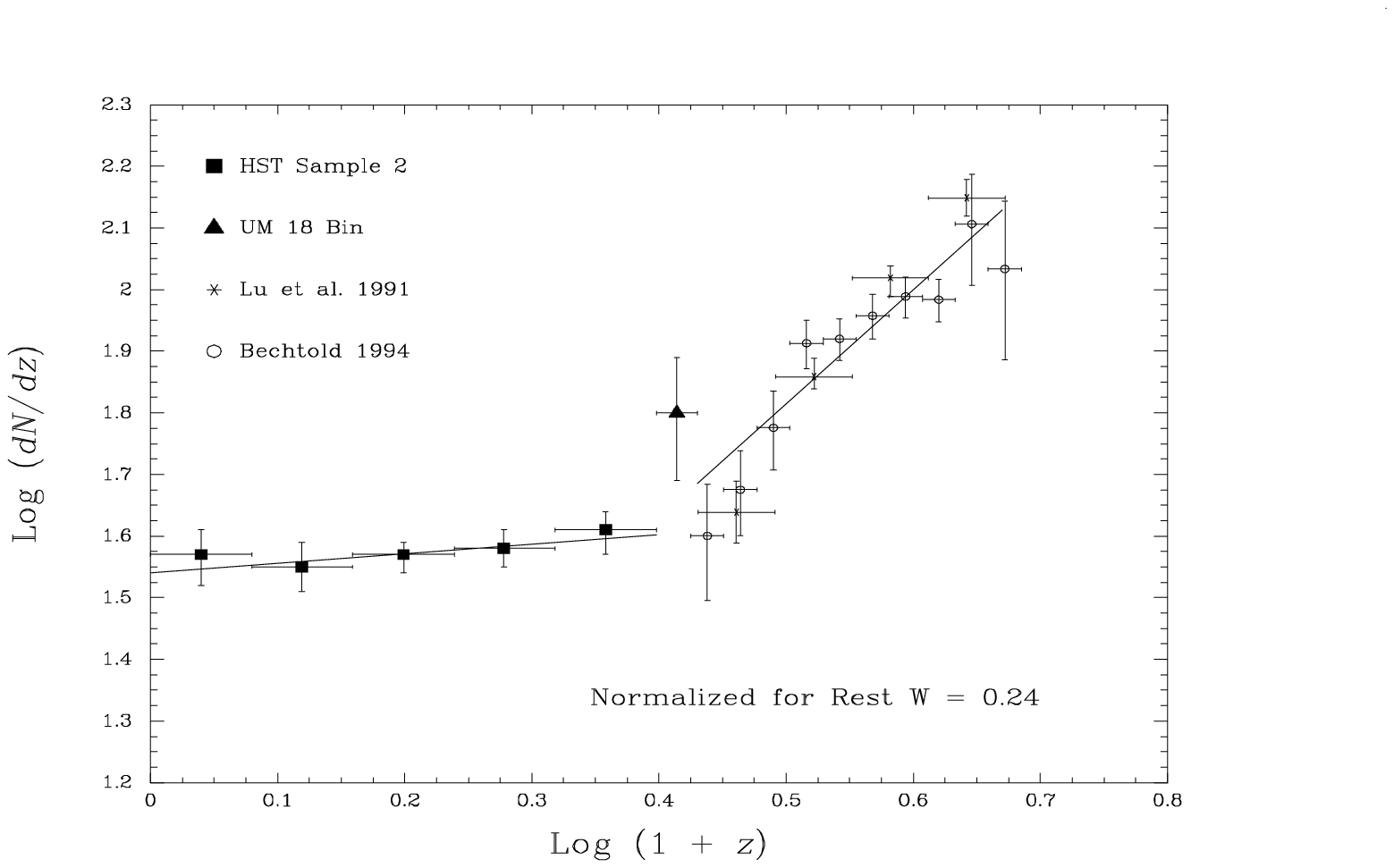,width=5in}}
\vglue-.1in
\caption[]{\footnotesize The log of the \Lya\ line density per unit redshift for
sample~2, evaluated at the midpoint of five bins equally spaced in
$\log (1+z)$ from $z=0$ to $z=1.5$ (solid squares), and for the \Lya\
lines in UM~18 in the redshift interval from 1.5 to 1.7
(triangle). The line through the solid squares is the best fit power
law for sample~2. The lighter symbols are taken from groundbased
surveys of \cite{lu91} (*'s) and \cite{bechtold94} (open circles). The
normalization for the groundbased data has been increased by a factor
$exp(0.36-0.24)/0.276$ to adjust the normalization from 0.36~to
the 0.24~\AA~used for the FOS data.  The line through the groundbased
data has a slope of 1.85, the best fit found by \cite{bechtold94}.
}
\end{figure}

\begin{figure}[t]
\centerline{\psfig{figure=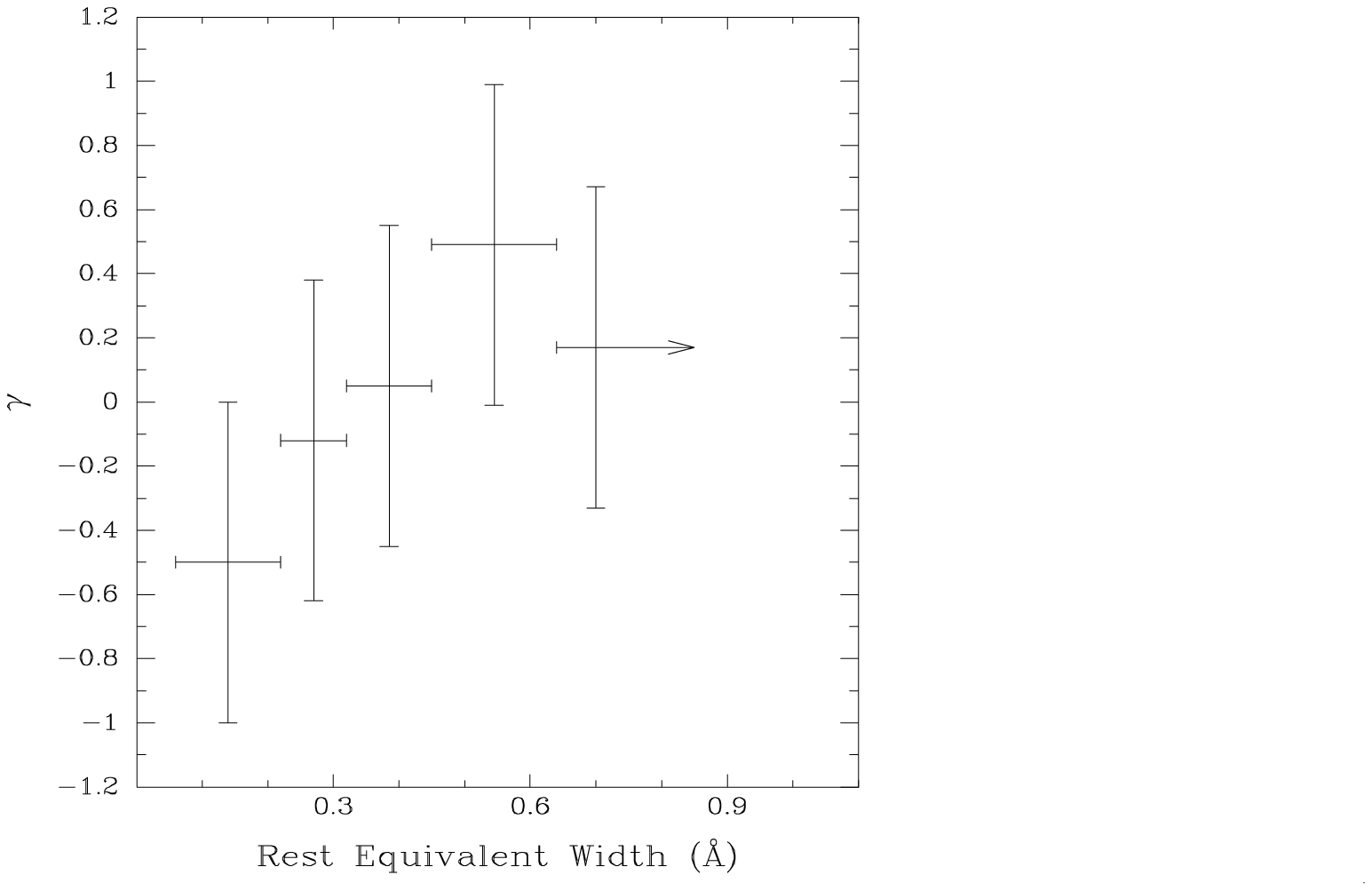,width=3.5in}}
\caption[]{\footnotesize The dependence of $\gamma$ upon rest equivalent width for
the lines in sample~2. The error bars are estimates which incorporate
both the formal fitting errors and the uncertainties associated with
the assumed value of $w^*$ in carrying out the fits. See the text for
further discussion.
}
\end{figure}


\begin{thebibliography}{}

\bibitem[Aldcroft {et~al.} 1997]{aldcroft97}
Aldcroft, T., Bechtold, J., \& Foltz, C. 1997, in
Mass Ejection from AGN, ASP Conference Series, 128, ed
N. Arav, I. Shlosman, R. Weymann p. 25 

\bibitem[Bahcall 1979]{bahcall79}
Bahcall, J. N., 1979, IAU Colloq. 54, Scientific Research With The
Space Telescope, ed. M.~S.~Longair \& J.~W.~Warner (Washington: GPO),
215

\bibitem[Bahcall \& Spitzer 1969]{bahcall69}
Bahcall, J. N., \& Spitzer, L. 1969, ApJ, 156, L63


\bibitem[Bahcall {et~al.} 1993]{bahcall93} 
Bahcall,
J. N., {et~al.} 1993, \apjs, 87, 1 (CAT1) 

\bibitem[Bahcall {et~al.} 1996]{bahcall96}  Bahcall,~J.~N., 
{et~al.} 1996, \apj, 457, 19 (CAT2)  

\bibitem[Barlow {et~al.} 1997]{barlow97}
Barlow, T., Hamann, F., \& Sargent, W.L.W. 1997, in Mass Ejection from
AGN, ASP Conference Series, 128, ed N. Arav, I. Shlosman, R. Weymann p
13.

\bibitem[Bechtold  1994] {bechtold94}
Bechtold, J. 1994, ApJS, 91, 1 


\bibitem[Bergeron \& Boisse 1991]{berg91}
Bergeron, J., \& Boisse, P. 1991, A\&A, 243, 344

\bibitem[Chen {et~al.} 1998]{chen98}
Chen, H.-W., Lanzetta, K. M., Webb, J.K., \& Barcons, X. 1998,
ApJ, 498, 77

\bibitem[Cowie {et~al.} 1995]{cowie95}
Cowie, L., Songaila, A., Kim, T-S., \& Hu, E. 1995, AJ, 109, 1522

\bibitem[1998]{cowsong98}
Cowie, L., \& Songaila, A. 1998, submitted to Nature

\bibitem[]{dave98}Dav\'e, \etal\ 1998, to be submitted to ApJ 

\bibitem[Dobrzycki \& Bechtold 1998]{dobrbech97}
Dobrzycki, A. \& Bechtold, J. 1998, in Proceedings of the 13th IAP
Astrophysics Colloquium: Structure and Evolution of the Intergalactic
Medium From QSO Absorption Line Systems, ed.~P.~Petitjean \&
S. Charlot, (Editions Frontier\'eres: Paris), 390

\bibitem[Foltz {et~al.} 1988]{foltz88}
Foltz, C., Chaffee, F., Weymann, R., \& Anderson, S. 1988,
in  QSO Absorption Lines: Probing the Universe ed.~J. Blades,
D. Turnshek \& C. Norman, 53 

\bibitem[Hamann {et~al.} 1997]{hamann97}
Hamann, F., Barlow, T., Cohen, R, Junkkarinen, V., \& Burbidge, E.M. 1997,
in Mass Ejection from AGN, ASP Conference Series, 128, ed.~N. Arav, I.
Shlosman, R. Weymann, 19

\bibitem[Impey {et~al.} 1996]{impey96}
Impey, C., Petry, C. E., Malkan, M. A., \& Webb, W. 1996,
\apj, 463, 473  

\bibitem[Jannuzi {et~al.} 1996]{jannuzi96}{Jannuzi, B. T., {et~al.}
1996, ApJ, 470, L1}  

\bibitem[Jannuzi 1998]{jannuzi98}
Jannuzi, B.~T.~1998, in Proceedings of the 13th IAP Astrophysics
Colloquium: Structure and Evolution of the Intergalactic Medium From
QSO Absorption Line Systems, ed.~P.~Petitjean and S. Charlot,
(Editions Frontier\'eres: Paris),~93  


\bibitem[Jannuzi {et~al.} 1998a]{jannuzi98a} Jannuzi,
B.T., {et~al.} 1998a, in press, \apjs, 118 (CAT3) 

\bibitem[Jannuzi {et~al.} 1998b]{jannuzi98b} Jannuzi,
B.T.,  {et~al.} 1998b,  in preparation 


\bibitem[Kim {et~al.}~1997]{kim97}
Kim, T-S., Hu, E. M., Cowie, L. L., \& Songaila, A. 1997, AJ, 114, 1

\bibitem[Lu {et~al.} 1991]{lu91}
Lu, L., Wolfe, A. M., \& Turnshek, D. A. 1991, \apj, 367, 19

\bibitem[Lu {et~al.} 1998a]{lu98a}  Lu, L.,  {et
al.} 1998a, in preparation  

\bibitem[Lu {et~al.} 1998b]{lu98b}  Lu, L.,  {et
al.} 1998b, in preparation 

\bibitem[Mo \& Miralda--Escude 1996]{mo96}
Mo, H. J., \& Miralda--Escude, J. 1996, \apj, 469, 589

\bibitem[Morris {et~al.} 1993]{morris93} 
Morris, S. L., {et~al.} 1993, \apj, 419, 524

\bibitem[Murdoch {et~al.} 1986]{murdoch86}  
Murdoch, H. {et~al.} 1986, \apj, 309, 19 

\bibitem[Press {et~al.} 1993]{press93}
Press, W. H., Rybicki, G. B., \& Schneider, D.~P.~1993, \apj, 414, 64 

\bibitem[Press {et~al.} 1986]{numrec}
Press, W. H., {et~al.} 1986, Numerical Recipes (Cambridge: Cambridge
University Press)

\bibitem[Riediger {et~al.} 1998]{riedig98}
Riediger,  R., Petitjean, P., \& Mucket, J. 1998, A \& A, 329, 30

\bibitem[Sargent {et~al.} 1980]{sybt80}
Sargent, W.L.W., Young, P., Boksenberg, A. \& Tytler, D.~1980,
ApJS, 42, 41

\bibitem[Sargent {et~al.} 1998]{sargberg}
Sargent, W.L.W., {et~al.} 1998, in preparation

\bibitem[Schneider {et~al.} 1993]{schneider93}
Schneider, D. P., {et~al.} 1993, ApJS, 87, 45 (Paper~II)

\bibitem[Steidel {et~al.} 1994] {steid94}
Steidel, C., Dickinson, M., \& Persson, E. 1994, 
ApJL, 437, L78

\bibitem[Stengler--Larrea {et~al.} 1995]{steng95}
Stengler--Larrea, E. A., {et~al.} 1995, \apj, 444, 64

\bibitem[Storrie--Lombardi {et~al.} 1996]{storrie96}
Storrie--Lombardi, L.J., McMahon, R. G., Irwin, M.J., \& Hazard, C. 1996,
\apj, 468, 121. 

\bibitem[Turnshek {et~al.} 1998]{turnshek98}
Turnshek, D. A., {et~al.} 1998, in preparation 

\bibitem[Williger {et~al.} 1994]{williger94}
Williger, G.M., Baldwin, J.A., Carswell, R.F.,
Cooke, A.J., Hazard, C., Irwin, M.J., McMahon, R.G., \& 
Storrie--Lombardi, L.J. 1994, \apj, 428, 574.

\end{thebibliography}
\end{document}